\documentclass[a4paper,11pt]{article}
\pdfoutput=1 

\usepackage{jcappub} 

\usepackage{lmodern} 
\usepackage[T1]{fontenc} 

\usepackage{xfrac} 
\usepackage[mathscr]{euscript} 
\usepackage{mathtools} 
\usepackage{soul} 
\usepackage{hyperref} 
\usepackage{siunitx}
\usepackage{mathabx} 
\usepackage{xspace} 
\usepackage{isotope}
\usepackage{booktabs} 

\usepackage{subcaption}
\usepackage{afterpage} 

\usepackage{environ}
\NewEnviron{mla}{%
	\begin{equation}\begin{split}
	\BODY
	\end{split}\end{equation}
}
\NewEnviron{mla*}{%
	\begin{equation*}\begin{split}
	\BODY
	\end{split}\end{equation*}
}

\newcommand{\nl}{\\&\qquad}

\DeclarePairedDelimiter\abs{\lvert}{\rvert}
\DeclarePairedDelimiter\norm{\lVert}{\rVert}

\newcommand{\eref}[1]{eq.\,(\ref{#1})}
\newcommand{\sref}[1]{section~\ref{#1}}
\newcommand{\aref}[1]{appendix~\ref{#1}}
\newcommand{\rcite}[2][]{ref.\,\cite[#1]{#2}}

\newcommand{\figref}[1]{figure~\ref{#1}}

\newcommand{\infrac}[2]{{\sfrac{#1}{#2}}}
\newcommand{\lfrac}[2]{{{#1}/\,{#2}}}

\newcommand{\ex}{\mathrm{e}}
\newcommand{\pp}{\pi}
\newcommand{\dd}{\mathrm{d}}

\renewcommand{\vec}[1]{\mathbf{#1}}

\renewcommand{\d}[2]{\frac{\dd #1}{\mathrm{\dd} #2}}

\newcommand{\ld}[2]{\lfrac{\dd #1}{\!\dd #2}}
\newcommand{\hyph}{-}

\newcommand{\dash}{~--- }

\newcommand{\dmwel}{\textsc{DMWEL}\xspace}

\newcommand{\aul}{A}
\newcommand{\baul}{\bar{\aul}}
\newcommand{\haul}{\hat{\aul}}

\newcommand{\ffo}{f_\text{fo}}
\newcommand{\ffoj}{f_{\text{fo},j}}
\newcommand{\jmax}{j_\text{max}}

\newcommand{\refequant}{\SI{20}{eV}}
\newcommand{\refe}[1][E_1]{\left(\frac{#1}{\refequant}\right)}

\newcommand{\refalpha}{\left(\frac{\alpha_1}{\num{e-3}}\right)}
\newcommand{\lrefalpha}{\left(\lfrac{\alpha_1}{\num{e-3}}\right)}

\newcommand{\pann}{p_\text{ann}}

\newcommand{\gss}{g_{\star\text{s}}}
\newcommand{\gssz}{g_{\star\text{s}0}}

\title{Dark matter with excitable levels}

\author{Andrew J. Wren}

\emailAdd{andrew.wren@ntlworld.com}

\abstract{This paper explores the possibility that dark matter particles are objects with a set of excitable levels, as for a string or drum.  A change in excitation is associated with the absorption or emission of a photon, and the probability of such a change is dependent on the ambient photon energy density.  Particle mass comes entirely from excitations. Such dark matter with excitable levels, or \dmwel, particles have a distribution of masses which freezes out to its present\hyph{}day form during the early universe.  In the particular model considered, consistency with Planck CMB anisotropy observations implies a cautious lower bound on the energy of the first excitable level of $\num{2}$~to~$\SI{200}{eV}$ for a set of reference parameters. While laboratory detection looks to require a considerable increase in the power of controlled light\hyph{}sources, it might be possible in the near future to detect or constrain \dmwel further through improved anisotropy observations, closer determination of the CMB's maximum deviation from a blackbody shape, or from gravitational probes of the dark matter power spectrum.  For certain parameters near to the boundary of the region allowed by CMB anisotropies, \dmwel dark matter is cold, but with a cut\hyph{}off halo mass many orders of magnitude greater than that typical for WIMP cold dark matter.}

\arxivnumber{1912.11453}

\begin{document}
\maketitle
\flushbottom

\section{Introduction}
	\label{sec:intro}

Following the principle of ``leaving no stone unturned'' in looking for dark matter~\cite{BertoneTait}, this paper proposes a new dark matter particle, based on the well\hyph{}known physical phenomenon of an object having a set of vibrational levels, as for a drum or a string.  Atomic electronic transitions suggest how changes in occupation of such levels may be associated with the absorption or emission of photons. This is used as motivation for a model of dark matter particles which are in principle detectable, but have properties consistent with current observational data because the transitions are suppressed sufficiently to avoid being obvious in the present day.  The model adopted here makes transition probabilities dependent on the ambient photon energy density\dash and so, in the cosmological context, on the photon temperature\dash the physical idea being that this energy influences the vibrational ``tension'' in the dark matter particle.  This model is described as \emph{dark matter with excitable levels} (\dmwel). 

As far as the author is able to determine, similar models have not previously been proposed.  A different notion of excitable levels occurs in models of composite dark matter, such as dark atoms ~\cite{2010JCAP...05..021K,2013PhRvD..87j3515C,2014PhRvD..89l1302C,2018PhRvD..97l3018G}, a dark sector analogy of an atom with electron\hyph{}like levels including, unlike in this paper's model, the possibility of ``ionisation''.  Dark atom transitions are associated with the emission or absorption, not of photons, but of radiation which is dark to avoid easy present\hyph{}day detectability.  Dark atoms' astrophysical detection primarily involves observing interactions or effects on the evolution of structure~\cite{2010JCAP...05..021K,2013PhRvD..87j3515C,2018PhRvD..97l3018G}, but can involve positing small dark sector\hyph{}photon interactions~\cite{2014PhRvD..89l1302C}. 

The \dmwel model is motivated by broad analogy with a drum or string; its only link with conventional standard model extensions is to borrow the string theoretic idea~\cite[see, for example,][]{polchinski_2000,becker_becker,dine_2007,zwiebach_2007} that mass arises solely from the energy of the excited levels.  That assumption enables the \dmwel model to predict particle mass from basic underlying parameters, and hence, assuming all dark matter is \dmwel, to derive its number density. Note that proposed models of dark matter using string theory have recently been reviewed in \rcite{2018JHEP...09..130A}, with the best tests  suggested to be via direct
detection experiments and direct collider searches.

This paper's \dmwel model leads to consideration of several cosmological routes for detection, relating to the CMB, nuclide abundances, and minimum halo sizes.  There may also be future prospects for laboratory detection, but these appear more remote.\\

The paper is arranged as follows. Section~\ref{sec:the-dmwel-model} describes in more detail the model of dark matter with excitable levels. Section~\ref{sec:dmwel-mass} considers the characteristics of \dmwel particles, including mass and number density. Section~\ref{sec:cosmological-constraints} considers cosmological constraints on \dmwel parameters, covering in successive subsections, CMB anisotropies, CMB energy density and spectrum, post\hyph{}BBN photodissociation of nuclei, and kinetic decoupling. Section~\ref{sec:detection-of-dmwel} considers whether \dmwel might be detected in the laboratory. Section~\ref{sec:conclusion} concludes with a discussion.  Appendices address the validity of the excitation equation (\ref{sec:validity-of-the-equation-erefeqf-for-excitation-fractions-f}), and set out details relating to \dmwel's photodissociation of nuclei (\ref{sec:photodissociation-of-light-nuclei}).

Unless otherwise indicated, all physical constants and quantities are taken from \rcite{2018PhRvD..98c0001T}. 
\emph{Mathematica} and Python calculations used in this paper, including to draw the figures, can be found in \rcite{num}.

\section{The model for dark matter with excitable levels}
	\label{sec:the-dmwel-model}

This section develops the \dmwel model outlined in the introduction, choosing a particular reference model from the very wide range of possibilities.  For simplicity, the energy levels will be taken to be fermionic in nature, being either empty or occupied.  Photon absorption is associated with a level going from empty to occupied, whilst photon emission is associated with going from occupied to empty.  For a string or drum, there are two transverse modes of a given energy\dash for a string, these are in the two orthogonal directions transverse to the string, and, for a drum, these are transverse vibrations directed along two orthogonal directions in the drum skin.  Motivated by this, \dmwel particles have a pair of levels of each given energy.  This also accounts for photon polarisations, the levels in a pair being associated with orthogonal polarisations.  

To detail the model further, assume that the photon background can be considered as a thermal bath, in the sense that the \dmwel particles have no significant influence on the photon temperature (this assumption is checked in \aref{sec:validity-of-the-equation-erefeqf-for-excitation-fractions-f} for the particular model outlined below).  Also assume that the photon spectrum does not significantly deviate from a pure blackbody spectrum (this will be checked in \sref{sec:cmb-energy-density-and-spectrum}).  Consider the emission/absorption probabilities for a given single level from the pair with energy $E.$ Emission and absorption probabilities can be given by a parameter $A,$ the Einstein coefficient representing the probability of a spontaneous emission~\cite[see, for example,][]{2002physics...2029H}. By a standard ``detailed balance'' argument, there are two other Einstein coefficients, associated with absorption, $B_\uparrow,$ and stimulated emission, $B_\downarrow,$ equal in value and proportional to $A.$ With the current conventions, they are related by 
\begin{equation}	\label{eq:Einstein-coeffs}
B_\uparrow
=
B_\downarrow
=
\frac {2 \pp^2\hbar^3 c^3\ }{E^3}\, \aul
.
\end{equation}
The next step is to derive an equation governing the probability that the level is occupied, its \emph{excitation fraction}, $f.$ To ensure this is a straightforward differential equation, the \dmwel particle must have a sufficiently large mass, so that (1) its velocity is non\hyph{}relativistic, with no need to account for photons being Doppler\hyph{}shifted, and (2) its mass is much larger than the energy of a single transitioning level.  These conditions are explored further in \sref{sec:simplifying-assumptions}, and their validity is confirmed for a wide range of parameters in \aref{sec:validity-of-the-equation-erefeqf-for-excitation-fractions-f}. Assuming that these conditions are satisfied, the energy of the absorbed or emitted photon is equal to the level energy $E,$ giving the excitation equation governing $f$ as
\begin{equation}	\label{eq:ems-abs}
\d{f}{t}
=
B_\uparrow  \, \rho_{\gamma,E}(T)  \left(1-f\right)
-
\left[
\aul
+
B_\downarrow \, \rho_{\gamma,E}(T) 
\right]
f
,
\end{equation}
where
\begin{equation}
\rho_{\gamma,E}(T)
=
\frac{E^3}{2\pp^2\hbar^3 c^3\left(\ex^{\infrac{E}{T}}-1\right)}
\end{equation}
is the unipolar spectral energy density, that is the differential rate of change of the energy density with respect to the photon energy, for a particular choice  of photon polarisation. 

The requirement for emission and absorption to depend on the ambient photon energy density suggests further setting
\begin{equation} \label{eq:baul-haul}
A
\equiv
\baul\,\rho_{\gamma}
\stackrel{\text{eq}\stackrel{\text{m}}{.}}{\equiv}
\haul \,T^4
,
\end{equation}
for constants $\baul$ and $\haul,$ where the second equivalence holds only if the ambient photons are in thermodynamic equilibrium, in which case their  unipolar energy density, totalled over all photon energies, is given by
$
\rho_\gamma(T)
=
\lfrac{\pp^2 \, T^4}{30\hbar^3 c^3} 
.
$
Using eqs.~\eqref{eq:Einstein-coeffs} and~\eqref{eq:baul-haul}, the excitation equation, \eref{eq:ems-abs}, then becomes
\begin{equation} \label{eq:time-deriv}
\d{f}{t}
=
\haul \, T^4 
\left(\frac{
	1
	-
	\left[\ex^{\infrac{E}{T}} +1\right]	f
	}{\ex^{\infrac{E}{T}}-1}\right)
,
\end{equation}
where $T$ is the photon temperature if the photon gas is in thermodynamic equilibrium, or, in other cases, an appropriate effective ``excitation temperature''.  For practical purposes, it will usually be better to introduce a dimensionless parameter $\alpha\equiv\lfrac{\haul\, E^4}{H(E)},$ which implies that
\begin{equation} \label{eq:alpha-j-1}
\haul
=
\SI{9.5e-19}{eV^{-4}.s^{-1}} \left(\frac{\alpha}{\num{e-3}}\right)
\left(\frac{E}{\refequant}\right)^{-4}
\left(\frac{H(E)}{H(\refequant)}\right)
,
\end{equation}
where the $\si{eV^{-4}.s^{-1}}$ quantity was calculated using the Friedman equation~\cite{num}.  To provide an initial condition for \eref{eq:time-deriv}, assume that at early times the \dmwel level is in thermodynamic equilibrium with the photon bath\dash and so for high temperatures $f(T)=f_\text{eq}(T)\equiv\lfrac{1}{(1+\ex^{\lfrac{E}{T}})}.$  As will be illustrated in \sref{sec:simplifying-assumptions} and \figref{fig:f}, this leads to a freeze\hyph{}out scenario, in which, as $T$ declines through $T=E,$ the level\hyph{}photon interaction becomes weaker, and the excitation fraction $f(T)$ ``freezes out'', becoming appreciably greater than $f_\text{eq}(T),$ and eventually asymptoting to some effectively constant freeze\hyph{}out value $\ffo.$

It remains to define the set of levels for the \dmwel particles\dash the choice of rule for this is pragmatic, picking a simple rule fairly arbitrarily.  Section~\ref{sec:intro} introduced the assumption that \dmwel mass comes only from excited levels, so if there were only a single pair of levels then a fraction $\left(1-f\right)^2$ of the particles would be massless, invalidating the conditions described after \eref{eq:Einstein-coeffs}. Instead, the reference model used in this paper assumes that there are many level pairs, which are labelled by integers $j=1,2,3,...,,\jmax,$ where $\jmax$ is taken to be very large and for practical purposes can be left unspecified (see the end of \aref{sec:validity-of-the-equation-erefeqf-for-excitation-fractions-f} for comment on how large $\jmax$ can be).  By assumption, the energy of an excitation at a $j$th level, $E_j,$ is proportional to $j^2,$ and all the Einstein coefficients, such as $\haul_j,$ are proportional to $j^{-2},$
\begin{equation}	\label{eq:e-a-rules}
E_j = E_1 \, j^2
\qquad\text{and, for example,}\qquad
\haul_j = \haul_1 \, j^{-2}
.
\end{equation}

To close this section, note that with this choice of  rules, the inverse $j$\hyph{}dependency of $E_j$ and $A_j$ leads to a $j$~invariant, but photon\hyph{}temperature dependent, tension which can be calculated as~\cite{num}
\begin{equation}	\label{eq:tension}
	\kappa(T)
	\equiv
	\frac{E_j A_j}{c}
	=
	\SI{1.0e-22}{eV.cm^{-1}} \left(\frac{T}{\refequant}\right)^4
	\refalpha \refe^{-3}\left(\frac{H(E_1)}{H(\refequant)}\right)
		,
\end{equation}
where it was assumed that light speed, $c,$ is the appropriate phase speed.  A $j$\hyph{} and $T$\hyph{}invariant area can also be constructed, and, in femtobarns, this is
\begin{equation}	\label{eq:cross-section}
\sigma
\equiv
\frac{E_j \baul_j}{c}
=
\SI{1.5e-2}{fb}
\refalpha \refe^{-3}\left(\frac{H(E_1)}{H(\refequant)}\right)
,
\end{equation}
which it is tempting to interpret as a cross section, perhaps associated with the dependency of emissions and absorptions on the ambient photon energy density. 

\section{\dmwel characteristics} 
	\label{sec:dmwel-mass}

This section explores characteristics of \dmwel particles.  The first subsection makes some simplifying assumptions, used only in \sref{sec:simplifying-assumptions} and \aref{sec:validity-of-the-equation-erefeqf-for-excitation-fractions-f}.  The second subsection explores the model further without the simplifying assumptions.
 
\subsection{Simplifying assumptions}
\label{sec:simplifying-assumptions}

For an indicative exploration of \dmwel characteristics it is useful to make the simplifying assumptions that interest is confined to the radiation\hyph{}dominated epoch and the number of relativistic degrees of freedom $g_\star$ does not depend on temperature. These assumptions strictly speaking only apply fully for, say, $\SI{20}{eV}\lesssim T\lesssim\SI{50}{keV},$ the lower limit ensuring radiation domination, and the upper limit constant $g_\star.$  However, they provide a useful, and not too inaccurate, qualitative picture outside those limits.

\begin{figure}
	\centering
	\includegraphics[width=0.8\linewidth]{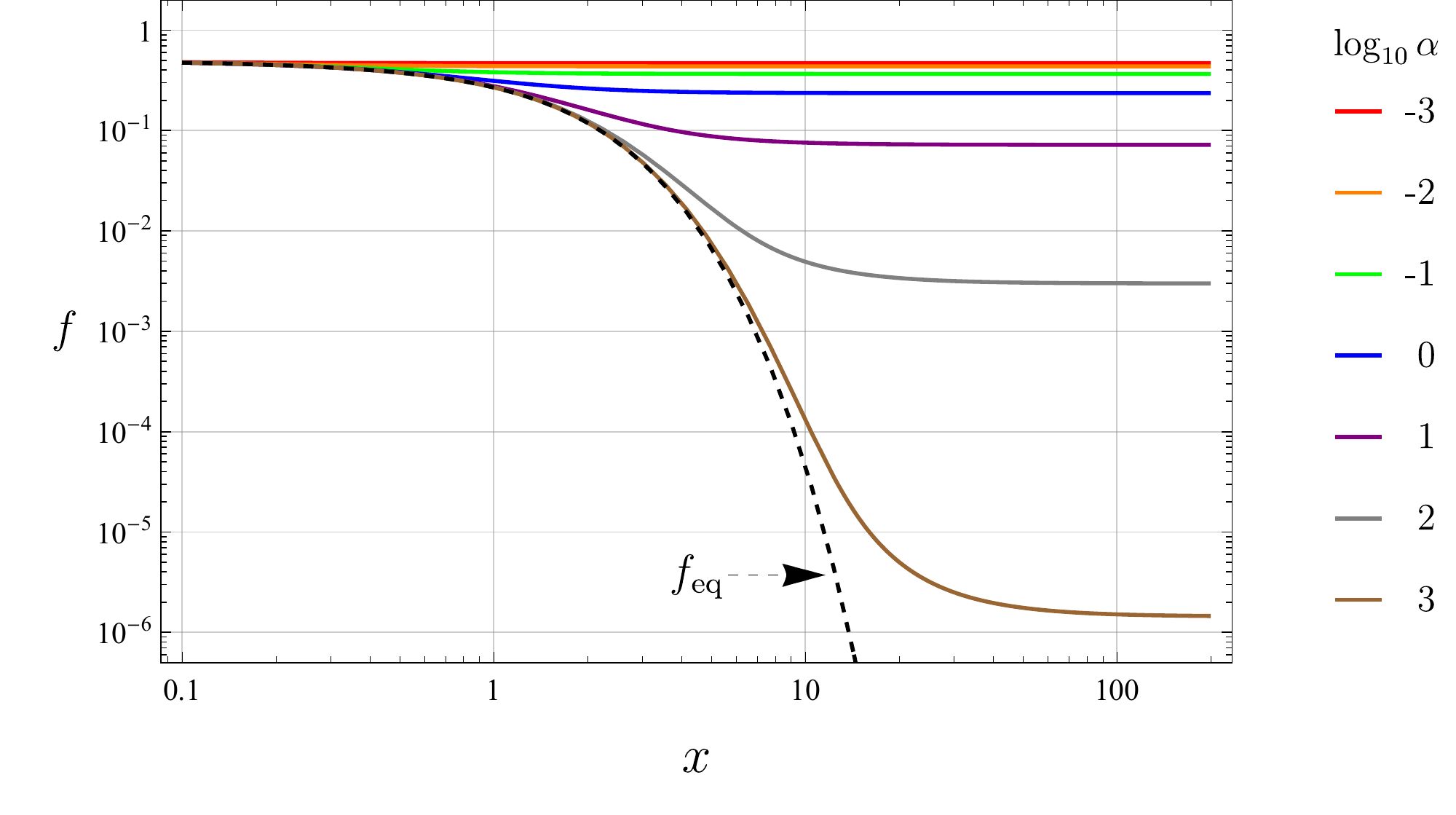}
	\caption{ 
		The freeze\hyph{}out process as $x$ varies, for selected of values of $\alpha.$ The function $f$ is the fraction of \dmwel particles for which the level is excited.  The dashed line, $f_\text{eq},$ with $f_\text{eq}(x)=\lfrac{1}{\left(1+\ex^x\right)},$ shows the excited fraction in equilibrium, which, assuming equilibrium, is independent of $\alpha.$  The figure makes the simplifying assumptions of \sref{sec:simplifying-assumptions}.}
	\label{fig:f}
\end{figure}

\begin{figure}
	\centering
	\includegraphics[width=0.7\linewidth]{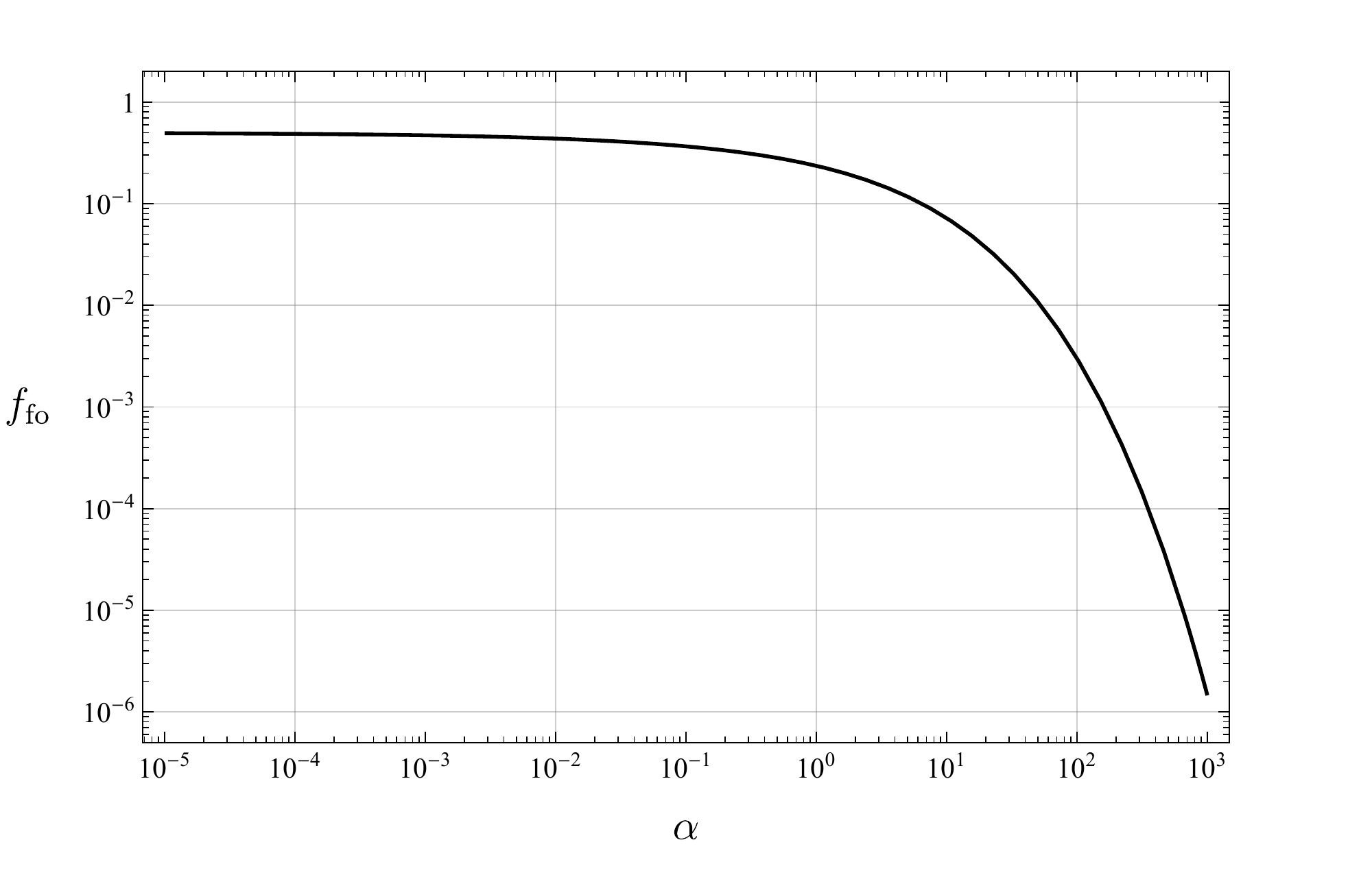}
	\caption{The excitation fraction at freeze\hyph{}out, as a function of $\alpha,$ with the simplifying assumptions of \sref{sec:simplifying-assumptions}.}
	\label{fig:ffo}
\end{figure}

With these assumptions, it is helpful to change the independent variable from $t$ to  $x=\lfrac{E}{T},$ noting that  $\ld{x}{t}= H(T) x=H(E)x^{-1},$ and that the excitation equation \eref{eq:time-deriv} becomes 
\begin{equation} \label{eq:f}
\d{f}{x}
=
\frac{\alpha\left[
	1 
	-
	\left(\ex^{x} +1\right)	f
	\right]}
{x^3\left(\ex^{x}-1\right)}
.
\end{equation}
Figure~\ref{fig:f} 
shows the resulting excitation fraction $f$ for different values of $\alpha$ and $x.$  As the temperature decreases, $x$ increases and the rate of change of $f$ tends to zero, leading to a freeze\hyph{}out value $\ffo$ of the excited fraction. Figure~\ref{fig:ffo} 
shows $\ffo$ as $\alpha$ varies.  Note also that the simplifying assumptions, with \eref{eq:e-a-rules} and the definition of $\alpha$ before \eref{eq:alpha-j-1}, imply $\alpha_j=\alpha_1\, j^2.$

\begin{figure}
	\centering
	\includegraphics[width=0.7\linewidth]{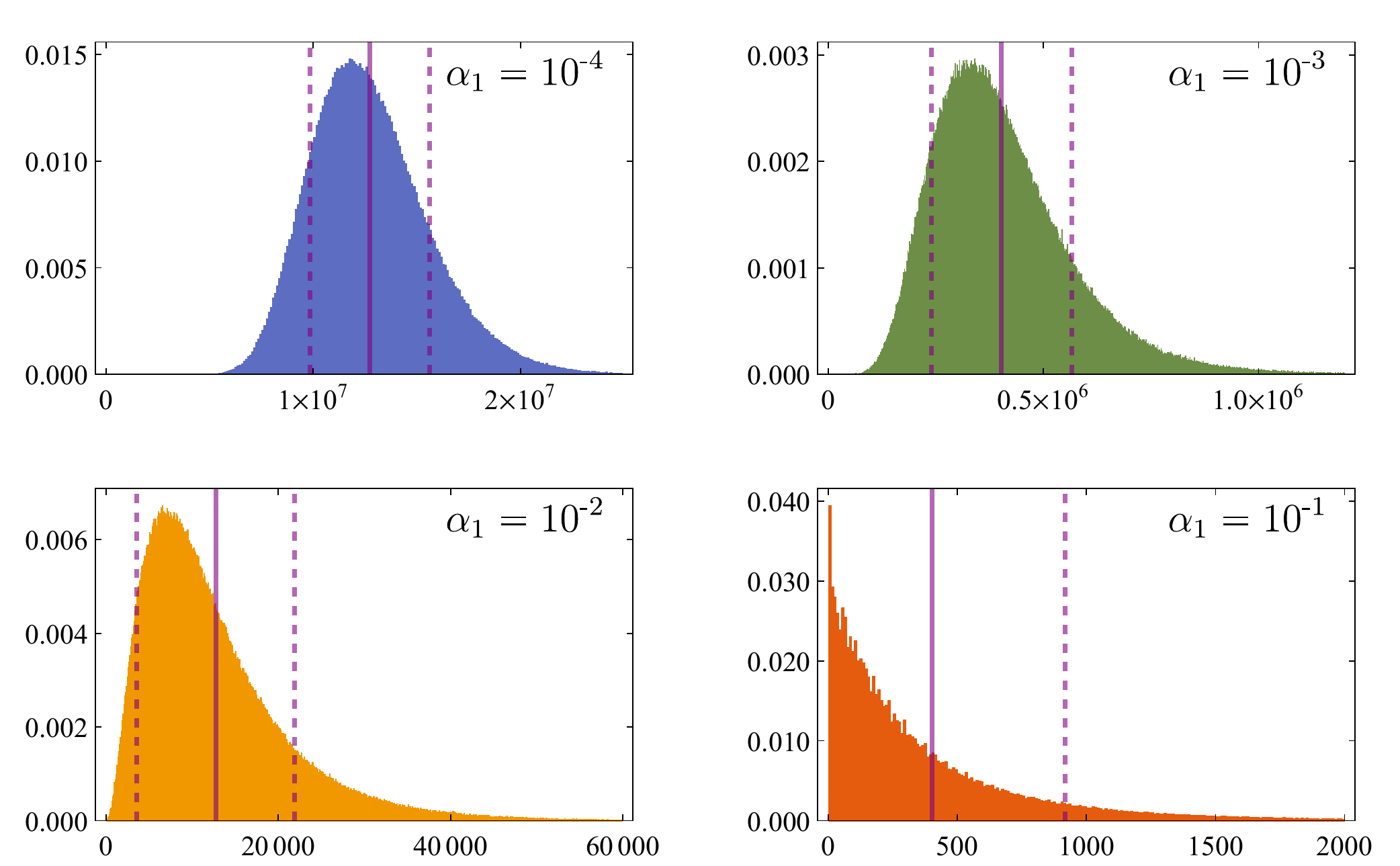}
	\caption{Freeze\hyph{}out mass distributions for indicated values of $\alpha_1.$ Horizontal axes represent mass in units of $E_1$, while vertical axes represent the probability of that mass. The solid vertical purple line corresponds to the mean mass, and the dashed purple lines to the limits of one standard deviation.  The figure makes the simplifying assumptions of \sref{sec:simplifying-assumptions}.  See the associated text on the calculation's invalidity for $\alpha_1=\num{e-1}.$}
	\label{fig:masshisto}
\end{figure}

Recall that the \dmwel particle mass comes exclusively from excitation energies.  Each of a \dmwel particle's levels has an independent probability of excitation, so the particle mass is not fixed, but follows a distribution given by 
\begin{equation}
m
=
2 E_1
\sum_j
j^2 F_j
,
\end{equation}
where $F_j=0,\infrac{1}{2},$ or $1,$ depending on whether neither, exactly one, or both the pair of $j$th levels are excited. For sufficiently large $x_1,$ the mass will freeze out to a distribution which is then  essentially fixed for subsequent times. Figure~\ref{fig:masshisto} shows freeze\hyph{}out mass distributions for a few selected values of $\alpha_1,$ based on a random sample of $\num{e6}$ particles for each $\alpha_1.$
The parameter $\alpha_1=\num{e-1}$ clearly requires some \dmwel particles to be highly relativistic, violating strongly conditions (1) and~(2) after \eref{eq:Einstein-coeffs}.

The simplifying assumptions are sufficient to resolve the applicability of conditions (1) and~(2) set out before \eref{eq:Einstein-coeffs}. Appendix~\ref{sec:validity-of-the-equation-erefeqf-for-excitation-fractions-f} shows that the non\hyph{}relativistic condition (1) is well\hyph{}satisfied for $\alpha_1\le\num{e-3}$ and more marginally satisfied for  $\alpha_1=\num{e-2}.$  Relativistic corrections will not affect the overall results of this paper, and will be disregarded.  Appendix~\ref{sec:validity-of-the-equation-erefeqf-for-excitation-fractions-f} also shows that the mass\hyph{}energy condition (2) is easily satisfied, at least for $\alpha_1\le\num{e-2}.$     For the remainder of this paper, it will be assumed that, $\alpha_1\le\num{e-2}$\dash  however, $\alpha_1>\num{e-2}$ is not necessarily infeasible, it is simply that, for such parameters, describing \dmwel's evolution requires more complicated equations than those used in this paper. 

\subsection{Without the simplifying assumptions}
\label{sec:without-the-simplifying-assumptions} 

Without the simplifying assumptions, it is useful to re\hyph{}write the excitation equation, \eref{eq:time-deriv}, in terms of $\alpha$ and the rate of change with respect to temperature, which gives
\begin{equation}	\label{eq:T-eqn}
\d{f}{T}
=
-
\frac{\alpha \,H(E)\, T^3 }{E^4 H(T)}
\left(1+\frac{1}{3}\d{\ln \gss}{\ln T}\right)
\left(\frac{
	1
	-
	\left[1+\ex^{\infrac{E}{T}}\right]	f
}{\ex^{\infrac{E}{T}}-1}\right)
,
\end{equation} 
because
\begin{equation}
\d{T}{t}=-H(T)T\left(1+\frac{1}{3}\d{\ln \gss}{\ln T}\right)^{-1}
,
\end{equation}
where $\gss(T)$ is the number of relativistic degrees of freedom for entropy density at temperature $T.$  The degrees of freedom, including the energy density\hyph{}related degrees of freedom which determine $H,$ are taken from data on the standard model in refs.~\cite{2016Galax...4...78H,2018PhRvD..98c0001T}.\footnote{The quark\hyph{}hadron transition is taken to occur at $\SI{160}{MeV},$ in the middle of the range $\num{150}$ to $\SI{170}{MeV}$ often quoted.}  Any additional high energy non\hyph{}standard model relativistic degrees of freedom would have to be very large in number to have a significant effect.

\begin{figure}
	\centering
	\begin{subfigure}{0.45\linewidth}
		\includegraphics[width=\linewidth]{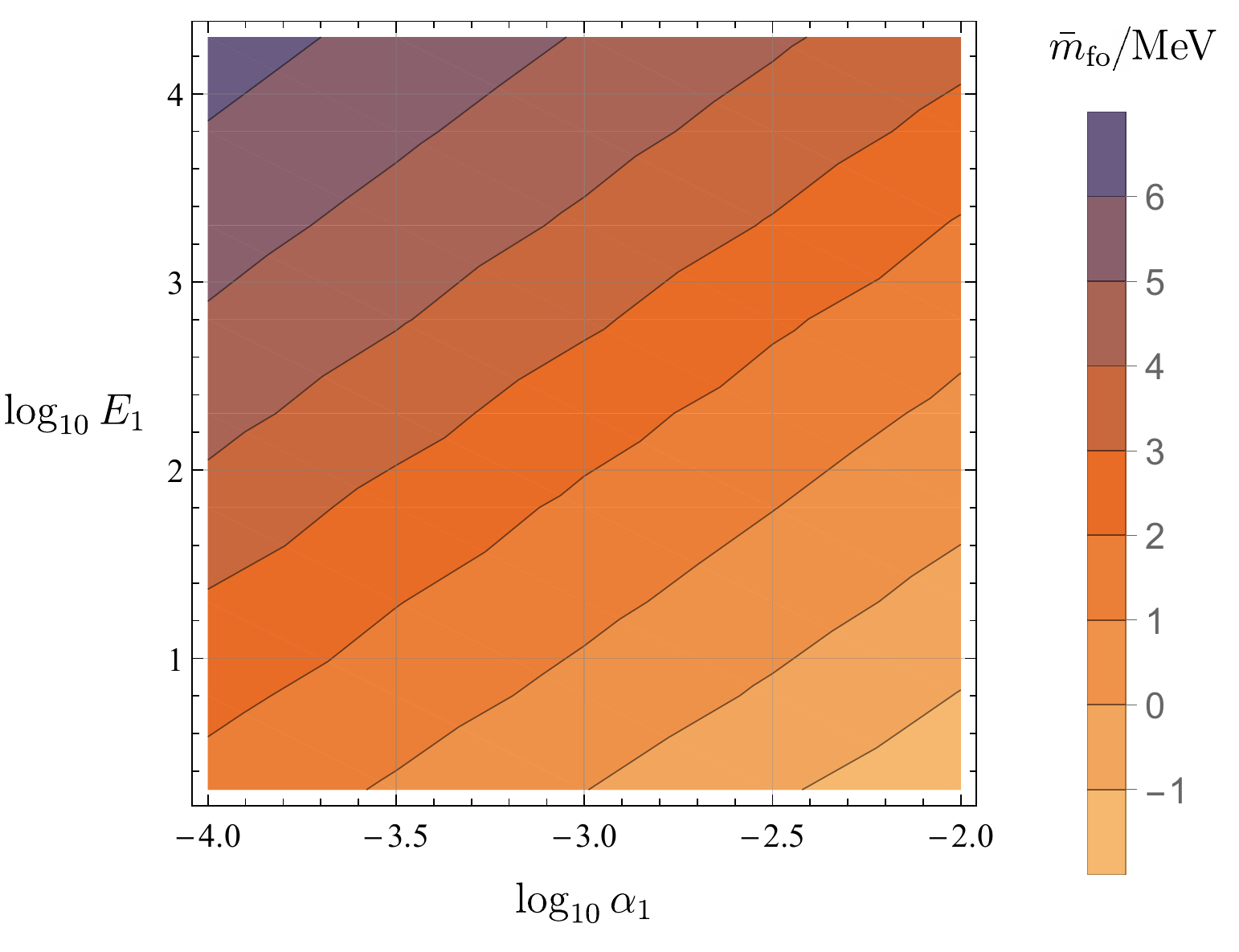}
		\label{fig:figmass}
	\end{subfigure}
	\qquad
	\begin{subfigure}{0.45\linewidth}
		\includegraphics[width=\linewidth]{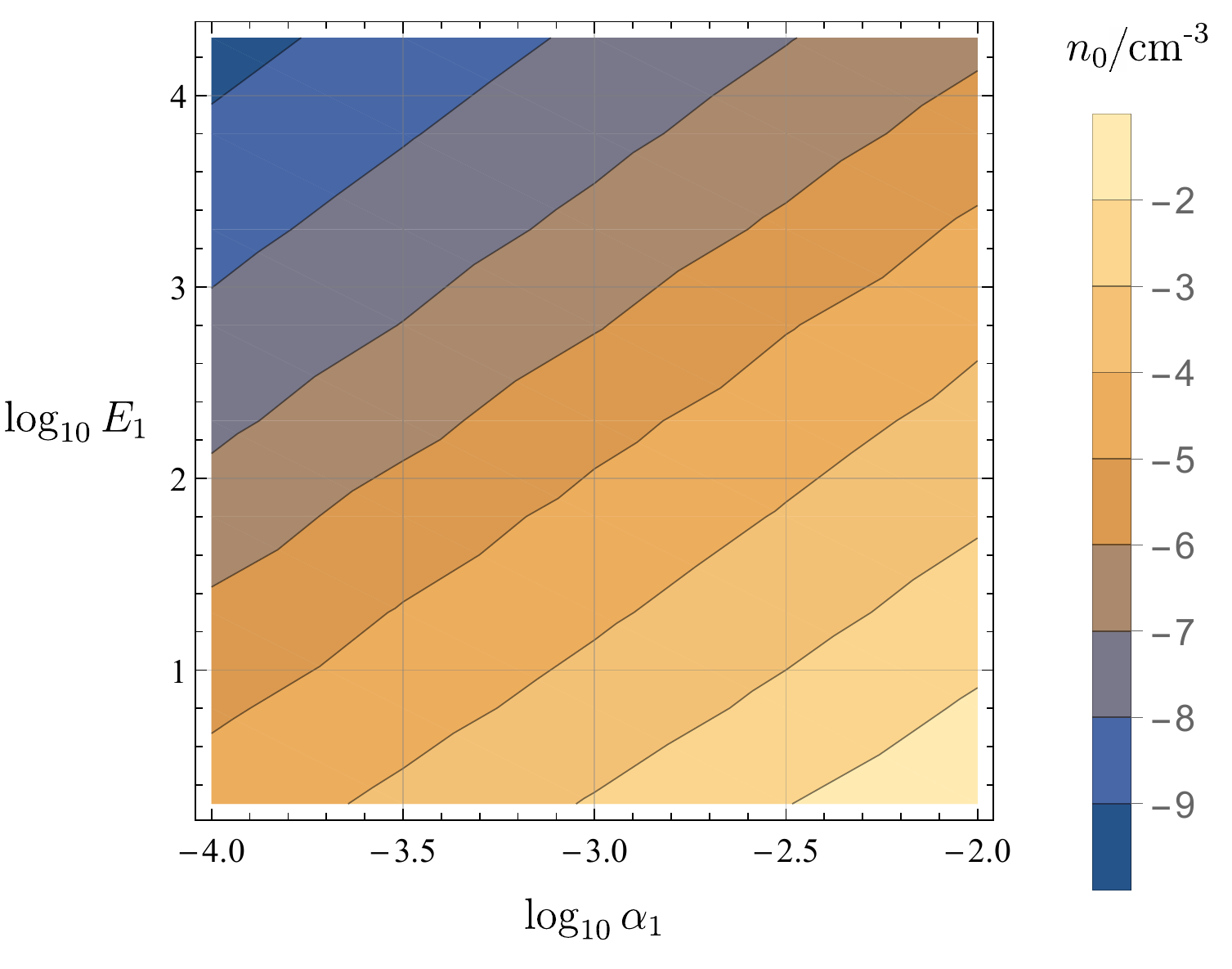}
		\label{fig:figndm}
	\end{subfigure}
	\caption{For selected \dmwel parameters,  mass after freeze\hyph{}out (\emph{left}) and  present\hyph{}day number density (\emph{right}).}
	 \label{fig:mfo-ndm0}
\end{figure}

Without the simplifying assumptions, from the definition of $\alpha$ before \eref{eq:alpha-j-1}, $\alpha_j$ is given by
\begin{equation}	\label{eq:alpha-j}
\alpha_j
=
\frac{\alpha_1\, j^6 H(E_1)}{H(E_j)}
.
\end{equation}
The mean mass over all \dmwel particles at a given temperature $T$ is then
\begin{equation}	\label{eq:mean-mass}
\bar{m}(T)
=
2 E_1
\sum_j
j^2 f(\alpha_j,E_1 j^2, T)
,
\end{equation}
and for low enough temperatures the mass has frozen out at an essentially constant value.  Assuming that \dmwel particles make up all dark matter, the present\hyph{}day \dmwel number density would then be $n_0=\lfrac{\rho_{\text{dm},0}}{\bar{m}_\text{fo}},$ the present\hyph{}day mass density divided by the mean freeze\hyph{}out mass.  The present\hyph{}day \dmwel particle mass and number density can be calculated numerically~\cite{num}, and are shown for selected parameters in \figref{fig:mfo-ndm0}.  The results for the number density enable verification in \aref{sec:validity-of-the-equation-erefeqf-for-excitation-fractions-f} that, as assumed at the start of \sref{sec:the-dmwel-model}, the photon background acts as a thermal bath, with \dmwel minimally affecting photon temperatures, except at very high temperatures of order $\SI{e5}{GeV}.$

\section{Cosmological constraints}
	\label{sec:cosmological-constraints}

This section considers cosmological constraints, arising from CMB anisotropies, the CMB energy density and spectrum, post\hyph{}BBN photodissociation of nuclei, and from kinetic decoupling.  Planck CMB anisotropy constraints emerge as the tightest.

\subsection[CMB anisotropies]{CMB anisotropies\footnote{Based on observations obtained with Planck (\url{http://www.esa.int/Planck}), an ESA science mission with instruments and contributions directly funded by ESA Member States, NASA, and Canada.}}
\label{sec:cmb-anisotropies}

Emission of photons in the early universe would have affected CMB anisotropies~\cite{2004PhRvD..70d3502C,2005PhRvD..72b3508P,2009PhRvD..80b3505G,2009PhRvD..80d3526S,2012PhRvD..85d3522F,2013PhRvD..88f3502G,2016A&A...594A..13P,2018arXiv180706209P}, so Planck observations, most recently updated in \rcite{2018arXiv180706209P}, can constrain \dmwel parameters. Such photons affect the intergalactic medium (IGM) through various ``channels'', of which ionisation of hydrogen atoms and heating of the IGM are significant for CMB anisotropies~\cite{2005PhRvD..72b3508P, 2012PhRvD..85d3522F,2013PhRvD..88f3502G}. Reference~\cite{2018arXiv180706209P} considered a model of dark matter self\hyph{}annihilation and derived a one\hyph{}sigma constraint on the self\hyph{}annihilation strength.  Here this constraint is used to derive related constraints for \dmwel dark matter's first energy level $E_1,$ for given $\alpha_1.$

If $\mathcal{D}_\text{c}$ is the energy per comoving unit volume deposited by \dmwel into channel $\text{c}$ at redshifts below $z,$ then, 
\begin{equation}	\label{eq:dmwel-cmb}
	\d{\mathcal{D}_\text{c}}{z}
=
n_0(\alpha_1,E_1)
\,
\sum_j
	2 f^\text{\tiny D}_{\text{c},\,j}(z) E_j\,\d{f_j}{z}
	,
\end{equation}
where $f^\text{\tiny D}_{\text{c},\,j}(z)$ is the power deposited in channel $\text{c}$ at redshift $z,$ having come originally from one of the $j$th levels, divided by the photon emission power of that level \emph{at} redshift $z.$ Similarly for dark matter self\hyph{}annihilation, the corresponding rate is given by~\cite{2016A&A...594A..13P}
\begin{equation}	\label{eq:ann}
\d{\mathcal{D}_\text{ann,\,c}}{z}
=
\frac{\chi_\text{c}\,\pann\,\rho_\text{dm,0}^2\left(1+z\right)^2}{H(z)}
,
\end{equation}
where $\chi_\text{c}$ is the fraction of the deposited energy that goes into channel $\text{c},$ and $\pann$ is a coefficient representing the self\hyph{}annihilation strength.  The most recent Planck results~\cite{2018arXiv180706209P} give a $95\%$\hyph{}confidence maximum of $\pann=\SI{3.2e-28}{cm^3.s^{-1}.GeV^{-1}},$ similar to, but slightly tighter than, the $\pann=\SI{3.4e-28}{cm^3.s^{-1}.GeV^{-1}}$ constraint from the Planck~2015 release~\cite{2016A&A...594A..13P}.\footnote{The parameter $\pann$  accounts for the total proportion of the energy which is deposited into the IGM~\cite[eq.~87]{2018arXiv180706209P}, which is why $\chi_c$ rather than $f^\text{\tiny D}_c$ appears in \eref{eq:ann}.}  The rates for \dmwel and self\hyph{}annihilation will turn out to have a broadly similar redshift dependency which will justify seeking values of $\left(\alpha_1,E_1\right)$ giving similar rates of energy deposit to those arising from the constraint value of  $\pann$ from \rcite{2018arXiv180706209P}.  This avoids doing a full Monte Carlo analysis of \dmwel emissions and cosmological parameters, as used in \rcite{2018arXiv180706209P} to find $p_\text{ann}$, providing instead a more direct comparison of the \dmwel and self-annihilation models. 

\begin{figure}
	\centering
	\includegraphics[width=0.9\linewidth]{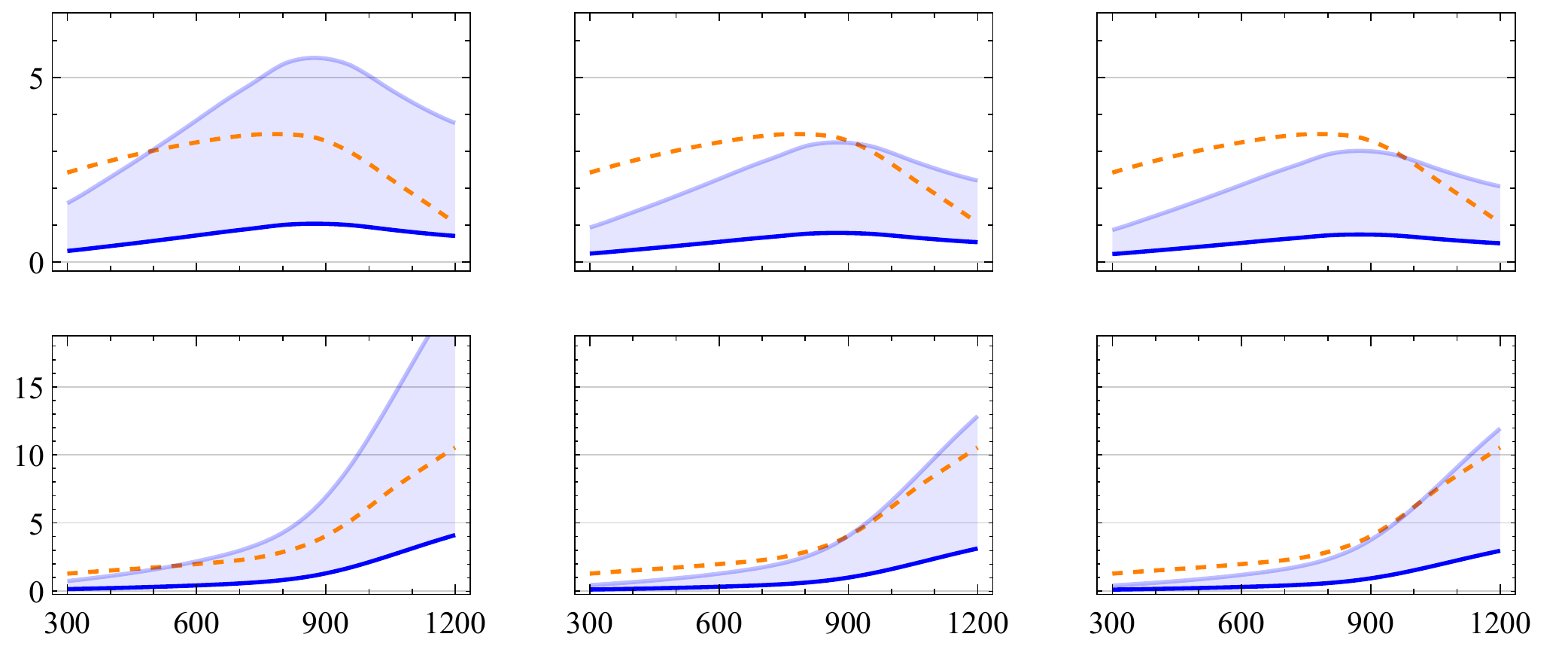}
	\caption{Energy deposit rates for \dmwel and self\hyph{}annihilating dark matter.  The plots show the  rate of energy deposit with respect to redshift per comoving $\SI{e12}{cm^{-3}},$ for redshifts $\num{300}\le z\le\num{1200}.$ The \emph{upper} (resp. \emph{lower}) row relates to deposit in the form of hydrogen ionisation (resp. plasma heating).  The columns represent $\alpha_1=\num{e-4},\num{e-3},\num{e-2}$ (\emph{left}, \emph{centre}, \emph{right}).  The orange dashed line is for self\hyph{}annihilating dark matter at the Planck  constraint,  $\pann=\SI{3.2e-28}{cm^3.s^{-1}.GeV^{-1}}$~\cite{2018arXiv180706209P}. The blue solid lines relate to \dmwel for the relevant $\alpha_1.$ The lower, thick, blue line is for $E_1=\refequant(\lfrac{\alpha_1}{\num{e-3}}),$  \eref{eq:CMB-min}'s  somewhat cautious constraint on $E_1;$ the upper, thin, blue line is for $E_1=\SI{10}{eV}(\lfrac{\alpha_1}{\num{e-3}}),$ showing that these lower values of $E_1$ could conflict with CMB anisotropy constraints.}
	\label{fig:CMB}
\end{figure}

The comparison of eqs.~\eqref{eq:dmwel-cmb} and~\eqref{eq:ann} can be facilitated by modifying slightly~\cite{num} the \texttt{DarkAges} package~\cite{2018JCAP...03..018S} from the \texttt{ExoCLASS} branch of \texttt{CLASS}~\cite{2011arXiv1104.2932L} to calculate the sum in \eref{eq:dmwel-cmb} for \dmwel, via \eref{eq:T-eqn}.  It can be found numerically that $\ld{\mathcal{D}_\text{c}}{z}$ is decreasing with $E_1$ over the parameter range used in \figref{fig:mfo-ndm0}, and at least down to $E_1=\SI{e-2}{eV}$~\cite{num}. In particular, results shown in \figref{fig:CMB}
indicate that, for $\num{e-4}\le\alpha_1\le\num{e-2},$ a cautious estimate of the minimum $E_1$ compatible with Planck observations is given by
\begin{equation}	\label{eq:CMB-min}
E_{1,\text{min}}
=
\refalpha\refequant
.
\end{equation}
In interpreting \figref{fig:CMB}, note that CMB anisotropies are most sensitive to self\hyph{}annihilation energy deposit in the range $600\lesssim z\lesssim\num{1000}$~\cite{2016A&A...594A..13P}.  \dmwel energy deposit's variation with redshift is similar to that for self\hyph{}annihilation, but slightly more biased to higher redshifts, suggesting that the greatest sensitivity to \dmwel emissions is in a redshift range similar to, but slightly higher than, $600\lesssim z\lesssim\num{1000}.$  This motivates the somewhat cautious choice of \eref{eq:CMB-min} as the constraint even for $\alpha_1=\num{e-3},\num{e-2},$ where the dashed annihilation line is above the upper \dmwel line for most of the redshift range shown.

\subsection{CMB energy density and spectrum}\label{sec:cmb-energy-density-and-spectrum}

The simplest cosmological constraint is the requirement to avoid there being a detectable background of \dmwel emissions.  The total present\hyph{}day \dmwel net emission energy density generated at temperatures $T_\text{d}\le T\le T_\text{u}$ is given by dividing the flux reaching a present\hyph{}day observer by the speed of light
\begin{mla}	\label{eq:rho-dm}
	\rho_{\gamma,\text{dm}}(T_\text{d},T_\text{u})
	&=
	-\frac{2}{c}\sum_j 
	\int_{T=T_\text{d}}^{T=T_\text{u}}\!\dd \chi\,4\pp\chi^2\, a^3\,
	n_0 a^{-3}\frac{a^2}{4\pp\chi^2} E_j \d{f_j}{t}
	\\
	&=
	n_0 a_0 T_0
	\int_{T_\text{d}}^{T_\text{u}}\!\frac{\dd T}{T}\,\left(\frac{\gssz}{\gss}\right)^\infrac{1}{3} 
	\d{\bar{m}}{T}
	,
\end{mla}
where $\chi$ is the comoving distance from the observer, $a(t)\dd\chi\equiv c\,\dd t,$ and $a$ is the scale factor.  From now on, it will be assumed that $a_0=1.$  For fixed $\alpha_1,T_\text{d}$ and $T_\text{u},$ $\rho_{\gamma,\text{dm}}(T_\text{d},T_\text{u})$ is decreasing with $E_1,$ because changes in mass occur at higher temperatures.

The earliest constraint on emissions comes from the constraints on photon density during Big Bang nucleosynthesis (BBN).  \dmwel emissions between that epoch and now contribute to the present\hyph{}day CMB, and so they imply a lower\hyph{}than\hyph{}standard photon temperature during BBN. In that context,  \rcite{2017IJMPE..2641007S} suggests that Planck observations constrain the CMB energy density at $T=\SI{10}{MeV}$ to be no more than around ten percent of the standard value. With \dmwel playing the part of dark matter, the relative CMB under\hyph{}density at $T=\SI{10}{MeV},$ compared with its standard density, is
$\lfrac{\rho_{\gamma,\text{dm}}(T_0,\SI{10}{MeV})}{\rho_\gamma(T_0)}.$
Calculations~\cite{num} then show that this quantity is less than $\num{e-4}$ for values of $\alpha_1$ and $E_1$ compatible with the anisotropy constraints of \eref{eq:CMB-min}, and so \dmwel parameters are not further constrained by effects during BBN.\\

Photon emissions at temperatures above around $\SI{500}{eV}$ are rapidly thermalised and affect only the blackbody temperature, not the CMB spectrum~\cite{2012MNRAS.419.1294C}.  However, later emissions may result in distortions to the CMB spectrum, which observations closely constrain to a blackbody spectrum~\cite{1996ApJ...473..576F,PDBook2002,2012MNRAS.419.1294C}.  Useful approximations for this effect are given in \rcite{2016MNRAS.460..227C}, taking account of the two possible types of primordial distortions, so\hyph{}called $\mu$\hyph{}type and $y$\hyph{}type distortions.  The simplest approximation considers separately the energy\hyph{}release in the $\mu$\hyph{}epoch, which is taken to be $T_\text{th}\equiv\SI{500}{eV}>T>\SI{10}{eV}\equiv T_{\mu\text{y}},$ and in the $y$\hyph{}epoch, $\SI{10}{eV}>T>\SI{0.26}{eV}\equiv T_\text{rec},$  the latter being the temperature at recombination. The respective spectral distortions are characterised via parameters which are given in the simplest approximation by~\cite{2012MNRAS.419.1294C,2016MNRAS.460..227C}
\begin{equation}	\label{eq:mu-y}
\mu
\equiv
\num{1.401}\frac{\Delta\rho_\gamma}{\rho_\gamma}\bigg|_\mu
=
\frac{\num{1.401}n_0 \,T_0}{\rho_\gamma(T_0)}\int^{T_\text{th}}_{T_{\mu\text{y}}}\frac{\dd T}{T}\d{\bar{m}}{T}
\end{equation}
and
\begin{equation}
y
\equiv
\frac{1}{4}\,\frac{\Delta\rho_\gamma}{\rho_\gamma}\bigg|_y
=
\frac{n_0\,T_0}{4\rho_\gamma(T_0)}\,\int^{T_{\mu\text{y}}}_{T_\text{rec}}\frac{\dd T}{T}\d{\bar{m}}{T}
.
\end{equation}	
The results~\cite{num} in table~\ref{tab:spectral-distortions} show that the relevant observational upper bounds from FIRAS~\cite{1996ApJ...473..576F} allow $E_1= E_{1,\text{min}}$ for $\num{e-4}\le\alpha_1\le\num{e-2},$ and so also all greater $E_1.$ In other words, the existing $\mu$ and $y$ constraints are much weaker than those from CMB anisotropies.  However, the $\mu$\hyph{}distortions of table~\ref{tab:spectral-distortions} are close to $5\sigma$\hyph{}detectable by the proposed PIXIE instrument~\cite[fig.~12 and the associated discusssion]{2011JCAP...07..025K}, and well within $5\sigma$\hyph{}detectability of the further proposed Super\hyph{}PIXIE~\cite{2019arXiv190901593C},  
while the $y$\hyph{}distortions are some way beyond both PIXIE and Super\hyph{}PIXIE's $1\sigma$\hyph{}sensitivity  ~\cite{2011JCAP...07..025K,2019arXiv190901593C}.\\
\begin{table}
	\centering 
	\begin{tabular}{ccS[table-format=-1e-1]S[table-format=-1e-1]S[table-format=-1e-1]cS[table-format=-1.1e-1]}
		\toprule
		&$\alpha_1$& \multicolumn{1}{c}{\quad\num{e-4}}& \multicolumn{1}{c}{\quad\num{e-3}} & \multicolumn{1}{c}{\quad\num{e-2}} && \text{FIRAS upper bound}  \\
			&$\lfrac{E_1}{\si{eV}}$& \multicolumn{1}{c}{\num{2}}& \multicolumn{1}{c}{\num{20}} & \multicolumn{1}{c}{\num{200}} && \\ \\
		$\mu$ &&
		 2e-7 & 1e-7 & 1e-7 &&9e-5 \\ 
		$y$ &&
		 5e-10 & 4e-10 & 3e-10 &&
		 1.5e-5  \\ 
		\bottomrule
	\end{tabular}
	\caption{The $\mu$ and $y$ parameters for selected $\alpha_1$ and the corresponding $E_1$ from the anisotropy constraints, displayed alongside the $(95\%$\hyph{}confidence) observational  upper bounds from FIRAS~\cite{1996ApJ...473..576F}.
	}
	\label{tab:spectral-distortions}
\end{table}

\begin{figure}
	\centering
	\hspace*{-2cm}\includegraphics[width=0.8\linewidth]{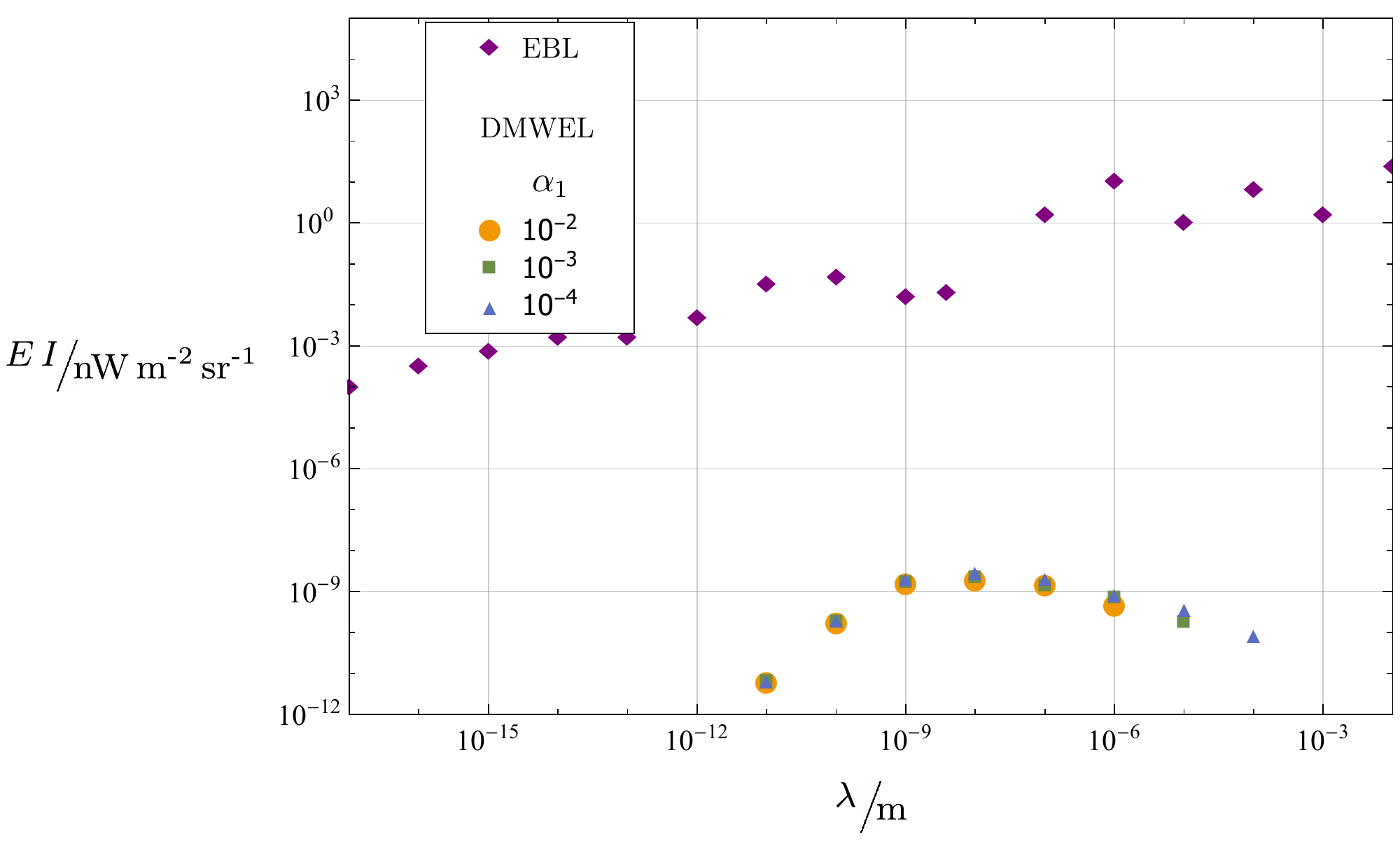}
	\caption{Comparing the spectral energy times the intensity of extragalactic background light (EBL)~\cite{2016RSOS....350555C} and \dmwel emissions for selected values of wavelength $\lambda$.  The \dmwel  parameters for the indicated $\alpha_1,$ with corresponding $E_1=\SI{20}{eV}\lrefalpha.$}
	\label{fig:figemissionsspectrum}
\end{figure}

It remains to quantify the spectrum of \dmwel emissions following recombination, via the \emph{intensity} $I,$ the incident radiation power per unit spectral energy per unit area per steradian.  The intensity at spectral energy $E$ associated with the $j$th pair of levels comes from emissions at temperature $T_j=\lfrac{E_j T_0}{E}$ and, from eqs.~\eqref{eq:rho-dm}, \eqref{eq:T-eqn} and~\eqref{eq:alpha-j}, is
\begin{equation}	
I_j(E)
=
\frac{c}{4\pp}\frac{2n_0 T_0 E_j}{T_j}
\,\d{f_j}{T}\Big|_{T_j}\d{T_j}{E}
=
\frac{c n_0 T_0^4\alpha_1  H(E_1) j^6 \ffoj}{2\pp E^4  H(T_j)}
,
\end{equation} 
while the total intensity at spectral energy $E$ comes from summing the over all $j$ for which $\lfrac{E}{E_1}\le j^2  \le \lfrac{T_\text{rec} E}{(T_0 E_1)}.$ The results~\cite{num} are shown in \figref{fig:figemissionsspectrum}, which compares the resulting intensity for the minimum anisotropy parameters of \eref{eq:CMB-min}, with that of the extragalactic background light (EBL)~\cite{2016RSOS....350555C}.  For relevant wavelengths near the \dmwel peak in \figref{fig:figemissionsspectrum}, larger values of $E_1$ give values of $E\,I$ which are lower.\footnote{This can be checked using \rcite{num}.}
Note that, for the \dmwel parameters plotted in the figure, the peak of \dmwel emissions strength is in a part of the ultraviolet where there is little observational data due to the absorption of this radiation by neutral hydrogen (this corresponds to the gap in the plotted EBL wavelengths around $\lambda=\num{e-8}$ to~$\SI{e-7}{m^{-1}}$)~\cite{2016RSOS....350555C}. Nonetheless, \figref{fig:figemissionsspectrum} clearly implies that post\hyph{}recombination \dmwel emissions are unobservable.

\subsection{Post\hyph{}BBN photodissociation of nuclei}
\label{sec:posthyphbbn-photodissociation-of-nuclei}

Photodissociation of light nuclei arises from gamma radiation with energies of at least around $E_\gamma=\SI{1.5}{MeV}$~\cite{2003PhRvD..67j3521C,2014PhRvD..90h3519I} trigger photodissociation at temperatures sufficiently low that these gamma rays no longer produce electron and positron pairs on CMB photons. The pair production threshold temperature is approximated by $T=\lfrac{m_\text{e}^2}{22 E_\gamma}$~\cite{1995ApJ...452..506K,2006PhRvD..74j3509J,2009NJPh...11j5028J,bertone_chapter}, which is a little less than $\SI{10}{keV}$ for $E_\gamma=\SI{1.5}{MeV},$ implying that for all relevant level energies, the excitation fractions have frozen out to $\ffoj,$ enabling the rightmost round round\hyph{}bracketed term in \eref{eq:T-eqn} to be replaced by $-\ffoj.$ Appendix~\ref{sec:photodissociation-of-light-nuclei} describes detailed photodissociation calculations based on a recently developed approach~\cite{2019JHEP...01..074F,2018JCAP...11..032H} for emissions of energy less that around $\SI{100}{MeV}.$  The appendix shows that photodissociation due to \dmwel is of negligible significance, at least for $10^{-4}\le\alpha_1\le 10^{-2}$ and the values of $E_1$ allowed by CMB anisotropies. Deuterium is the nuclide most sensitive to \dmwel emissions, and\ observational error would need to be reduced by more than ten  times in order for deuterium abundance to improve the constraints from CMB anisotropies.

\subsection{Kinetic decoupling}
\label{sec:kinetic-decoupling}

\begin{figure}
	\centering
	\includegraphics[width=0.7\linewidth]{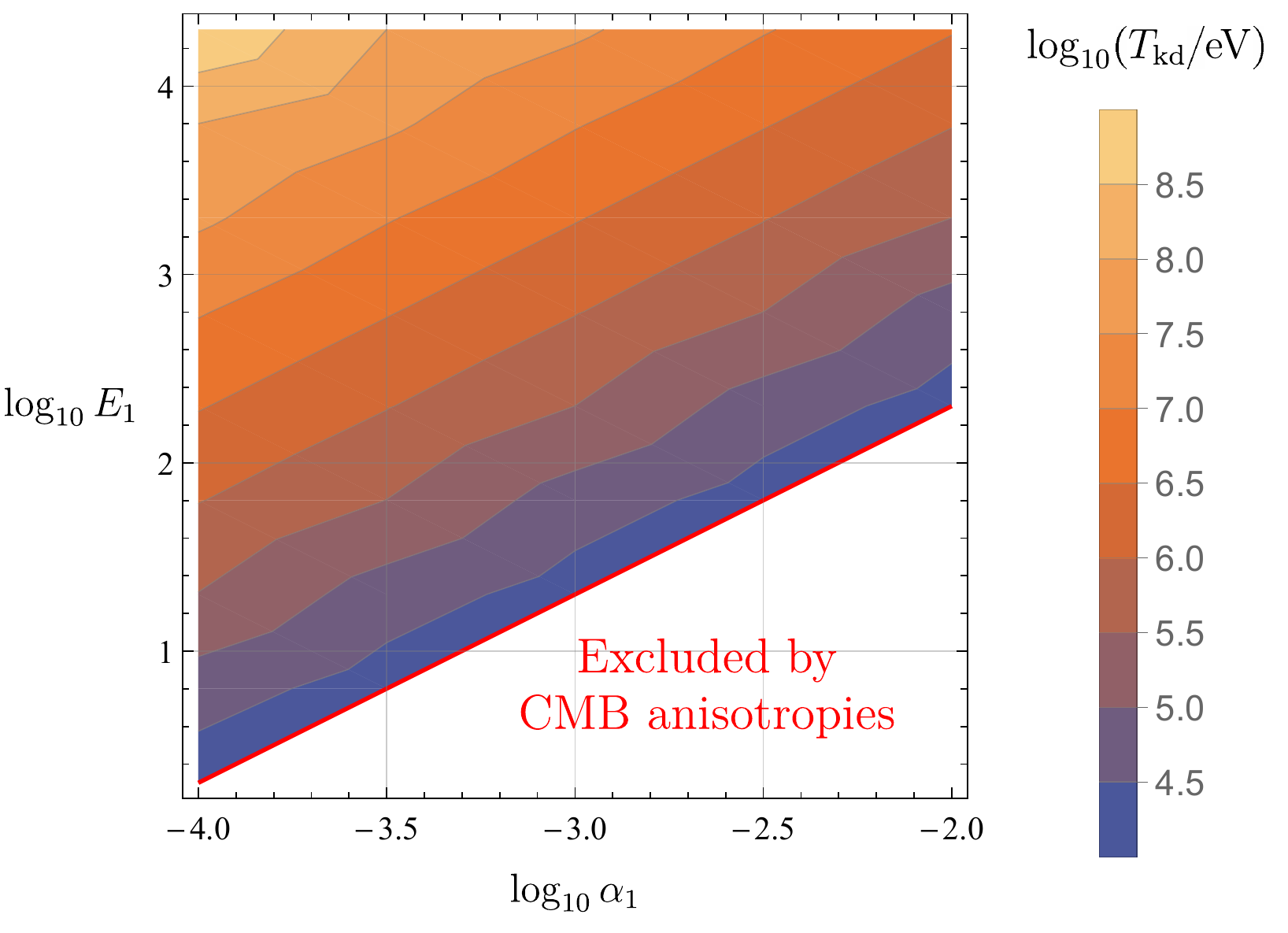}
	\caption{Kinetic decoupling temperatures for a range of \dmwel parameters.}
	\label{fig:figTkd}
\end{figure}

Kinetic decoupling\dash  which influences the matter power spectrum\dash  occurs when photon\hyph{}dark matter interactions are no longer frequent enough to maintain dark matter particles at momenta associated with the photon temperature, $p=\sqrt{2m T},$ within a Hubble time, $H^{-1}$~\cite{2005JCAP...08..003G,bringmann2009particle,cornell2012earthly}. From \eref{eq:time-deriv}, the rate per unit time for emissions and absorptions from the pair of levels $j$ is given by
\begin{equation}	\label{eq:e-a-rates}
r_{\text{e}\,j}
=
\frac{2\haul_1 j^{-2}\, T^4\ 
	\ex^{\lfrac{E_j}{T}} 	f(\alpha_j,E_j,T)
}{\ex^{\lfrac{E_j}{T}}-1}
\quad\text{and}\quad
r_{\text{a}\,j}
=
\frac{2\haul_1 j^{-2}\,T^4 \left(1-f(\alpha_j,E_j,T)\right)}{\ex^{\lfrac{E_j}{T}}-1}
\end{equation}
respectively, and in either case the size of the momentum change is the same, $\Delta p=E_j.$ The overall rate of variance in \dmwel momentum is then the sum over $j$ of $\left(r_{\text{e}\,j}+r_{\text{a}\,j}\right)E_j^2.$ From eqs.~\eqref{eq:e-a-rates} and~\eqref{eq:alpha-j}, the time period $t_\text{relax}$ over which that variance totals to the square of the typical kinetic equilibrium momentum is given by
\begin{equation}	\label{eq:kd}
t_\text{relax}\left(\frac{2\alpha_1  H(E_1) T^4}{E_1^2}\right)
\sum_j
j^2
\left(
\frac{1}{\ex^{\lfrac{E_j}{T}}-1}
+
f(\alpha_j,E_j,T)
\right)
=
2 m T
.
\end{equation} 
Kinetic decoupling occurs when the relaxation time becomes as long as the Hubble time, $t_\text{relax}=H(T)^{-1},$ so \eref{eq:kd} means that the kinetic decoupling temperature $T_\text{kd}$ satisfies
\begin{equation}	\label{eq:kd2}
\sum_j
\frac{2 j^2}{\ex^{\lfrac{E_j}{T_\text{kd}}}-1}
=
\frac{m(T_\text{kd})}{E_1}
\left(\frac{E_1^3 H(T_\text{kd})}{\alpha_1   T_\text{kd}^3 H(E_1)}-1\right)
,
\end{equation} 
and the resulting $T_\text{kd}$ are shown in \figref{fig:figTkd}.

\begin{figure}
	\centering
	\includegraphics[width=0.7\linewidth]{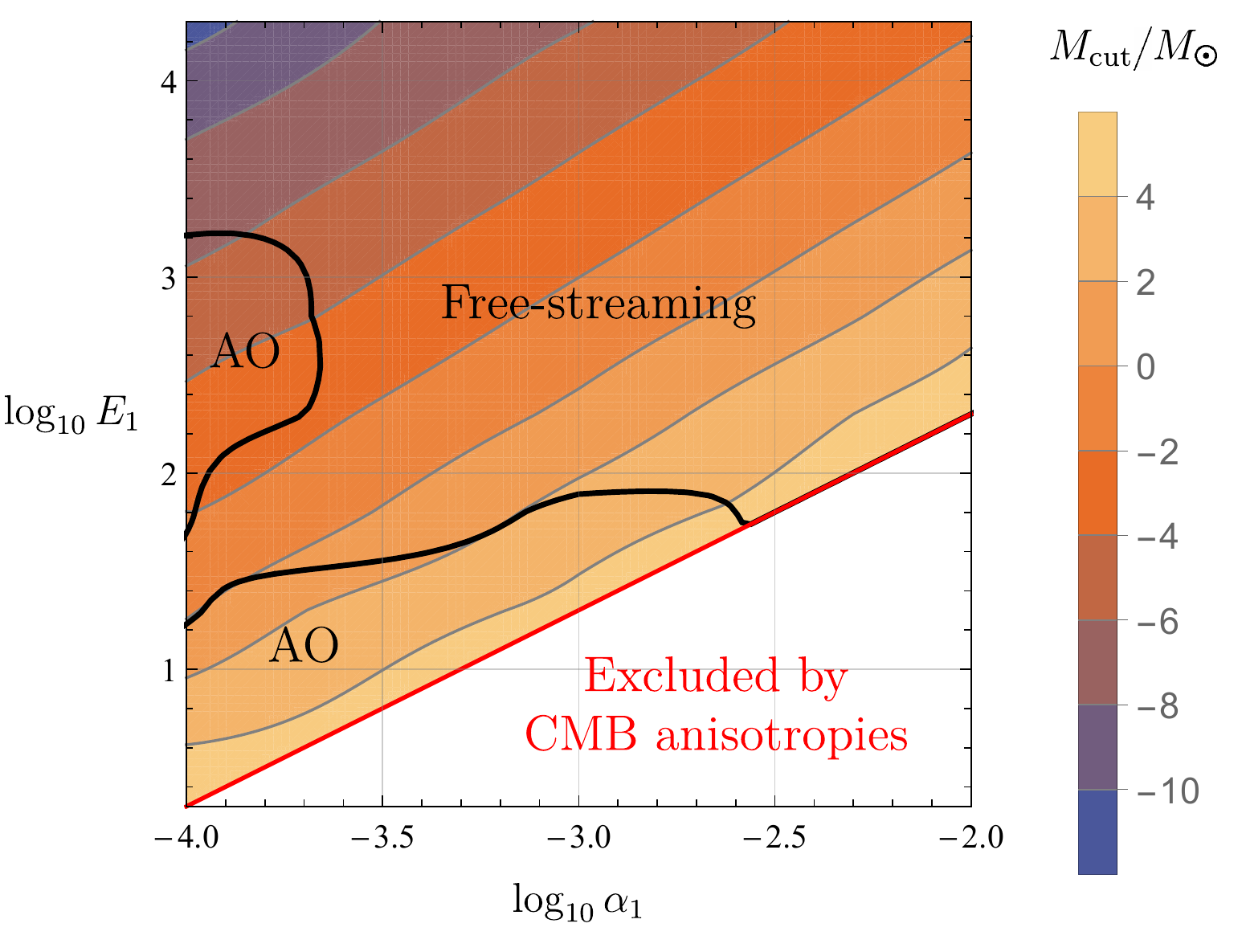}
	\caption{Cut\hyph{}off masses, $M_\text{cut},$ for a range of \dmwel parameters.  The black line divides  parameters where the cut\hyph{}off is determined by the acoustic oscillation (``AO'') scale, $k_\text{ao},$ of \eref{eq:ao-scale}, from those where it is determined by the free\hyph{}streaming scale, $k_\text{fs},$ of \eref{eq:fs-scale}.}
	\label{fig:figcutoff}
\end{figure}

There are three different scales associated with kinetic decoupling which each make a cut\hyph{}off in the power spectrum for dark matter perturbations. These scales are produced by acoustic oscillations, free\hyph{}streaming and damping, with comoving wave numbers $k_\text{ao}, k_\text{fs}$ and $k_\text{d}$ respectively. The dominant effect comes from the largest length scale of the three, which is the one with smallest comoving wave number. The acoustic oscillation scale~\cite{bringmann2009particle} is given by
\begin{equation}	\label{eq:ao-scale}
k_\text{ao}
=
\pp a_\text{kd} H_\text{kd}
,
\end{equation}
while the free\hyph{}streaming scale is~\cite{2005JCAP...08..003G} 
\begin{equation}	\label{eq:fs-scale}
k_\text{fs}
=
\left(\frac{m}{T_\text{kd}} \right)^\infrac{1}{2}
\frac{\left(\lfrac{a_\text{eq}}{a_\text{kd}}\right)}{\ln\left(\lfrac{4 a_\text{eq}}{a_\text{kd}}\right)}
\,k_\text{eq}
,
\end{equation}
where $k_\text{eq}$ is the comoving Hubble parameter and $a_\text{eq}$ the scale factor at radiation\hyph{}matter equality, both being taken from \rcite{2018arXiv180706209P}.\footnote{The mass explicitly appearing in \eref{eq:fs-scale} is the freeze\hyph{}out mass which affects the final free\hyph{}streaming, and is a little smaller than the mass at temperature $T_\text{kd}.$}  The damping scale is given by~\cite{2005JCAP...08..003G}
\begin{equation}	\label{eq:d}
k_\text{d}
=
\left(\frac{3}{2}\int_0^{t_\text{kd}}\!\frac{T\,t_\text{relax}}{m \,a^2 }\,\dd t\right)^{\!\!\infrac{-1}{2}}
, 
\end{equation} 
where $t_\text{kd}$ is the time of kinetic decoupling, and the integral is evaluated numerically, after changing the integration variable to $\ln T$~\cite{num}. The resulting mass cut\hyph{}off scale is then
\begin{equation}
M_\text{cut}
=
\frac{4\pp}{3}
\rho_{\text{dm},0}
\left(
\frac{\pp}{\min\left(k_\text{ao},k_\text{fs},k_\text{d}\right)}
\right)^{3}
,
\end{equation}
and this is shown for selected \dmwel parameters in \figref{fig:figcutoff}.  Over the parameter space displayed, the cut\hyph{}off mass comes from acoustic oscillations or free\hyph{}streaming.  The largest cut\hyph{}off masses occur at the CMB anisotropy constraints, being of order $\SI{e4}{M_\Sun}.$
For comparison, cut\hyph{}off masses for WIMPs are typically much smaller, ranging up to around $\SI{e2}{M_\Earth}\sim\SI{3e-4}{M_\Sun}$~\cite{2006PhRvL..97c1301P}.  Figure~\ref{fig:figcutoff}'s \dmwel cut\hyph{}off masses do not conflict with Lyman\hyph{}$\alpha$ forest constraints for the cut\hyph{}off at around $\num{1}h\,\si{Mpc^{-1}}$~\cite{viel2013warm,mcquinn2016evolution}, corresponding to $M_\text{cut}=\SI{e13}{M_\Sun}.$  Nor do they conflict with maximum cut\hyph{}off masses of \num{e7}~to \SI{e8}{M_\Sun} recently derived using quasar gravitational lensing~\cite{2020MNRAS.491.6077G} and from Gaia and Pan\hyph{}STARRS observations of stellar streams~\cite{2019arXiv191102663B}.

Possible approaches to observing cut\hyph{}off masses are discussed in refs.~\cite{bringmann2009particle,cornell2012earthly}.  Some  of these seek signals from dark matter annihilation boosted by halo densities, and so are not applicable to \dmwel.  Others involve the detection of dark matter halos through their gravitational effects, and are relevant. For example, it might be possible to confirm a lack of halos below a cut\hyph{}off mass through gravitational lensing~\cite{moustakas2009strong,chen2010gravitational} or through Doppler effects on millisecond pulsar timing
measurements, which could potentially be probed via the SKA and follow\hyph{}up observations~\cite{baghram2011prospects}.

\section{Laboratory detection}
\label{sec:detection-of-dmwel}

Detecting \dmwel in  astrophysical contexts would be difficult, requiring a hot, dense electromagnetic environment, in combination with the possibility of detecting emission or absorption of photons by the (fairly sparse) \dmwel particles, against that bright, hot background.  This suggests laboratory detection as potentially a better prospect, because detection could be optimised in a controlled environment.  

The present\hyph{}day local dark matter density at the Earth is estimated as $\rho_{\text{dm},0}\,\delta_{\text{dm}}^{\,\Earth} = \SI{0.5}{GeV.cm^{-3}}$~\cite{2018PhRvD..98c0001T,2015ApJ...814...13M}.  The imagined experiment, motivated by \rcite{2018EPJC...78..512D}, uses a powerful photon source, such as an X\hyph{}ray free\hyph{}electron  laser (XFEL)~\cite{2017RPPh...80k5901S}, to generate a high enough energy density for spontaneous \dmwel emissions to be observed.  In a configuration with laser photons of energy $\SI{1680}{eV},$ the European XFEL~\cite{XFEL} can generate an average power of $P_\text{\tiny EX}=\SI{1.6e21}{eV.s^{-1}},$ which is distributed via $\num{27000}$ photon bunches per second~\cite{tschentscher2017photon}.

Looking for spontaneous emissions, from eqs.~\eqref{eq:baul-haul} and~\eqref{eq:time-deriv} there would be an emission rate for the $j$th pair of levels,
\begin{mla}
	\mathcal{E}_j
	&=
	 \frac{2 n_0\,\delta_{\text{dm}}^{\,\Earth}\,\baul_j\,P L\,\ffoj}{c} 
	\\
	&=
	j^{-2}
	\refalpha
	\left(\frac{n_0}{n_0|_{ \alpha_1=\num{e-3},\,E_1=\refequant}}\right)
	\left(\frac{\ffoj}{\num{0.5}}\right)	
	\left(\frac{P}{P_\text{\tiny EX}}\right)
	\left(\frac{L}{\SI{1}{m}}\right)
	\,\SI{5e-11}{yr^{-1}}
	,
\end{mla}
for average power $P,$ where $L$ is the length of  the beam surrounded by a cylindrical detector of \dmwel emissions.\footnote{Note that this does not depend on any assumption about the excitation temperature of the \dmwel levels, which was defined after \eref{eq:time-deriv}.}  In order to generate a few first level emissions per year, the power (times the detector length) needs to be increased by a factor of  $\num{e11},$ and, because $\sum_j j^{-2}\approx 1.6,$ emissions from higher levels make little difference.  Increasing the power compared with the European XFEL would involve one or both of increasing the energy in each photon bunch and increasing the proportion of the time during which XFEL photons are generated (assuming that a photon bunch typically has a duration of, say, $\SI{2e-14}{s}$~\cite{tschentscher2017starting}, photons are only being emitted by the European XFEL for $\num{5e-10}$ of its operational time).  Note that the coherence of the beam is not important, nor need the collimation of the beam be very tight, however it does need to be free of photons of energy $E_j,$ and the detection volume needs to be very well\hyph{}shielded, in order to avoid any background photons of energy~$E_j.$  In summary, this section suggests that it is not yet possible to detect \dmwel dark matter in a laboratory, and to do so would require a considerable increase in the power of controlled light\hyph{}sources.

\section{Discussion}\label{sec:conclusion}

This paper considered a form of dark matter with excitable levels, \dmwel, for which the emission and absorption rates are proportional to the ambient photon energy density.  Beyond this interaction with photons, there is no direct link with the standard model of particle physics, however, the model reflects familiar physical notions of energy levels and their excitation, adding an extra ingredient of dependence on the ambient environment, which can be associated, in eqs.~\eqref{eq:tension} and~\eqref{eq:cross-section}, with a physical tension and cross section.  

For the particular model adopted, the strongest parameter constraints come from Planck observations of CMB anisotropies~\cite{2018arXiv180706209P}. Better constraint, or detection, via the abundance of light elements, or through laboratory methods, does not appear particularly plausible in the near future\dash and constraint or detection of \dmwel by observing emissions since recombination seems impossible.  The Planck constraints could be improved, or \dmwel detected, via improved anisotropy observations, better observations of the CMB spectrum~\cite{2011JCAP...07..025K,2019arXiv190713195K,2019arXiv190901593C}, or by detecting gravitational effects from the cut\hyph{}off \dmwel would induce in the dark matter power spectrum~\cite{baghram2011prospects,moustakas2009strong,chen2010gravitational}.

\dmwel gives a viable model of dark matter, which can be tested by observation.  Appendix~\ref{sec:validity-of-the-equation-erefeqf-for-excitation-fractions-f} confirms that, with parameter $\alpha_1\le\num{e-2},$ \dmwel is cold dark matter, and \sref{sec:kinetic-decoupling} finds that, for parameters near to the boundary of the region allowed by CMB anisotropies, the associated dark matter power spectrum cut\hyph{}off mass is many orders of magnitude greater than that typical for WIMP cold dark matter.

As noted at the end of \aref{sec:validity-of-the-equation-erefeqf-for-excitation-fractions-f}, at very early times \dmwel with extremely large numbers of levels will be sufficiently massive that level\hyph{}occupation stores all or almost all the energy that later is in the  radiation background. It is tempting to associate this with an end to inflation somewhat different to the usual reheating scenario, with kinetically cold \dmwel particles storing all the subsequent radiation heat.  The implications of this could be explored further.

This paper has focused on a part of \dmwel parameter space where the Einstein coefficients are sufficiently low, corresponding to $\alpha_1\le\num{e-2},$  to ensure that \dmwel occupation fractions can be modelled via a differential equation, \eref{eq:time-deriv}.  In particular, this is associated with the \dmwel particles being sufficiently non\hyph{}relativistic even at the time in the early universe when the ratio of \dmwel mass over photon temperature is at a minimum.  It should be possible to develop an approach for calculating the evolution of \dmwel particles with parameter $\alpha_1>\num{e-2},$ based on \eref{eq:f-rel}, and explore the associated dark matter which, since the particle mass would follow a distribution at least roughly like that of \figref{fig:masshisto}~(bottom right), looks likely to be a mixture of dark matter of varying warmth.  It would also be possible to use \eref{eq:time-deriv} to explore lower $\alpha_1$ parameters than those considered in this paper, $\alpha_1<\num{e-4}$ allowing first energy levels $E_1$ lower than $\SI{2}{eV}.$

Alternative models along \dmwel lines could, of course, be constructed.  There are limitless possible choices of rules characterising the temperature dependence, set here in \eref{eq:baul-haul}, and the $j$~dependence of the energy and Einstein coefficients, see
\eref{eq:e-a-rules}.  It would also be possible, for example, to replace the assumption that the levels are fermionic (occupation number zero or one) and make them instead bosonic (occupation characterised by a non\hyph{}negative integer), or indeed adopt some other occupation rule.  Another variant would be to associate a level change with absorption or emission of a particle other than a photon.  Wanting the dark matter to be plausibly detectable might suggest this should be an easily detectable particle, such as an electron, instead of, say, a neutrino which might always pass undetected. 

Another approach would be to explore \dmwel\hyph{}like models for which level occupation freezes in, rather than, as in this paper, freezes out.  Instead of starting in equilibrium at high temperatures and undergoing net emission, dark matter would begin with levels unoccupied at high temperatures, and  absorb photons over time.  This would imply greater photon density at early times relative to the standard cosmological model, and it is perhaps conceivable that such a model could explain the difference between values of the Hubble parameter detected via  early universe and local universe methods~\cite{2016ApJ...826...56R,2016JCAP...10..019B,2016PhLB..762..462K,2018JCAP...09..025M,2019arXiv190401016A,2019PhRvL.122v1301P, 2019NatAs...3..891V}.

In conclusion, this paper considered a new type of dark matter model, developing an example that was found to be consistent with  observational data and potentially detectable in the near future. Referring back to the start of \sref{sec:intro}, this indicates new ``stones'' to turn over in looking for dark matter~\cite{BertoneTait}.\\ \\ \\

\appendix
\noindent\textbf{\large Appendices}
\section{Validity of the excitation equation}
	\label{sec:validity-of-the-equation-erefeqf-for-excitation-fractions-f}

As indicated after \eref{eq:Einstein-coeffs}, for the excitation equation \eref{eq:time-deriv} to be valid, the \dmwel particle must (1) be sufficiently non\hyph{}relativistic that the the Doppler effects on the total and spectral photon energy density are not significant. It is also necessary (2) that its mass is much greater than the level energy, in order to avoid the energy of absorbed or emitted photons being too different from the corresponding \dmwel level energy. This appendix explores the validity of these conditions, together the validity of the assumption, set out at the start of \sref{sec:the-dmwel-model}, that \dmwel particles do not significantly affect the photon temperature. 

This appendix focuses on $\alpha_1= \num{e-2},\num{e-3}$ and $\num{e-4}.$ Figure~\ref{fig:masshisto} shows  that for $\alpha_1=\num{e-1},$ at least some \dmwel particles are relativistic, and so the excitation equation is not strictly valid.  The results for $\alpha_1=\num{e-4}$ will show that the excitation equation is clearly valid for lesser values of $\alpha_1.$

\begin{figure}
	\centering
	\includegraphics[width=0.7\linewidth]{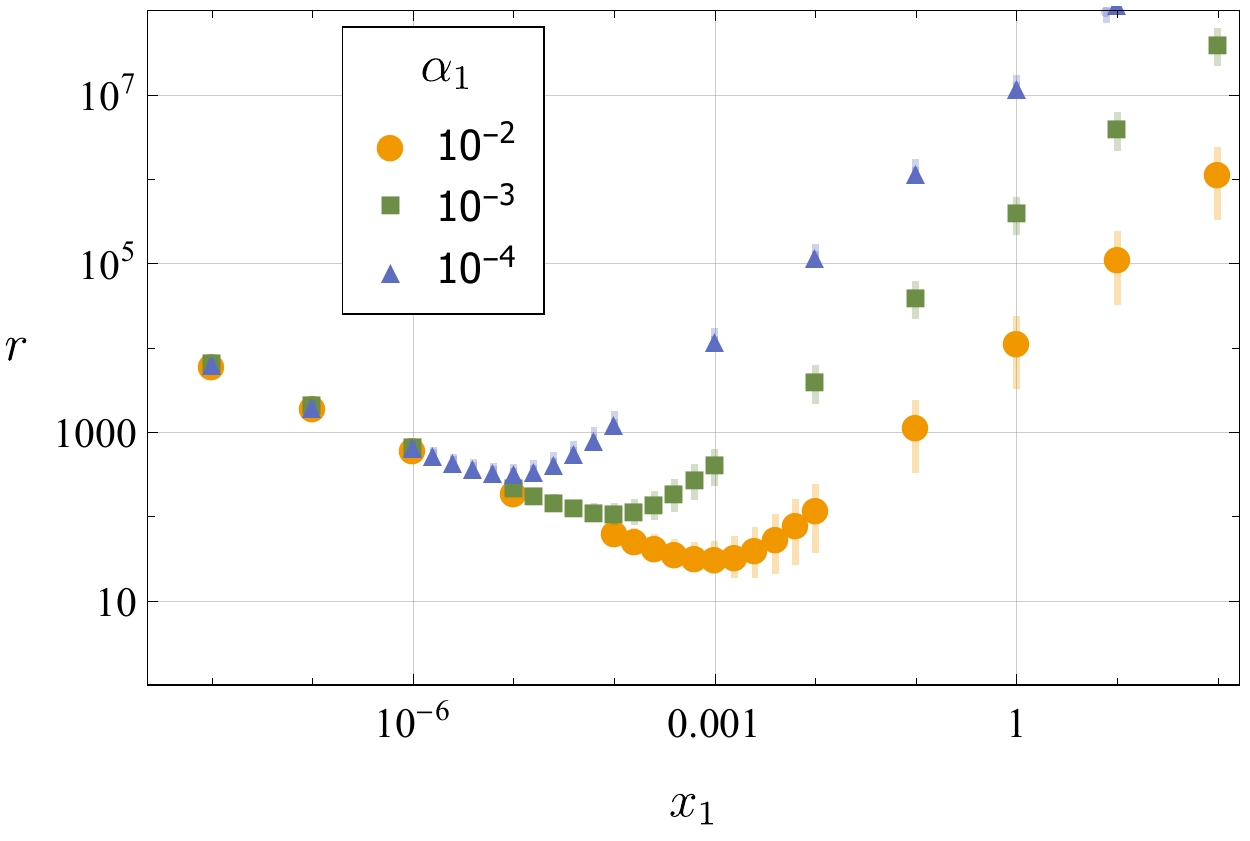}
	\caption{The ratio $r$ of mass divided by temperature for the indicated values of $\alpha_1$ and $x_1.$ The discs, triangles and squares show mean values, with the associated vertical lines indicating one standard deviation.}
	\label{fig:fignonrel}
\end{figure}

Figure~\ref{fig:fignonrel} 
 shows mass over temperature ratios $r=\lfrac{m}{T}$ for \dmwel particles with $\alpha_1= \num{e-2},\num{e-3}$ and $\num{e-4},$ indicating, for each $x_1,$ the mean ratio and one\hyph{}standard\hyph{}deviation range of $r.$  The lower end of those ranges has a minimum of around $r=\num{20},\num{90}$ and $\num{300}$ for $\alpha_1= \num{e-2},\num{e-3},\num{e-4}$ respectively. For convenience, in this appendix choose units so the speed of light $c=1.$ Then the velocity distribution of  \dmwel particles can be found from the relativistic gamma factor $\gamma = \left(1-v^2\right)^\infrac{-1}{2},$ which follows a Maxwell\hyph{}J\"{u}ttner distribution given by
\begin{equation}	\label{eq:M-J}
J(\gamma)
=
\frac{r\, \gamma^2 v}{K_2(r)}\exp(-r\,\gamma)
,
\end{equation}
where $K_2$ is a Bessel K\hyph{}function.

A \dmwel particle with a velocity $\vec{v}$ relative to the photon background can be taken~\cite{1968Natur.219.1343B,1968PhRv..174.2168P,1968PhRv..176.1451H,2009EL.....8820004N} to have a spectral energy density with effective temperature given by 
\begin{equation}	\label{eq:eff-T}
T_\text{eff}
=
\frac{T\sqrt{1-v^2}}{1-v\cos(\theta)}
,
\end{equation}
where $v=\norm*{\vec{v}},$ and $\theta$ is the angle between $\vec{v}$ and the direction of the photon observed from the particle. Considering the excitation equation, \eref{eq:time-deriv}, this leads to a relativistic version of \eref{eq:f},
\begin{equation} \label{eq:f-rel}
\d{f}{x}
=
\frac{\alpha}{x^3}
\left[
	\left\langle
		\frac{\left(\lfrac{T_\text{eff}}{T}\right)^4}{\ex^{\left(\lfrac{T}{T_\text{eff}}\right)x}-1}
	\right\rangle
	-
	\left\langle
	\frac{\left(\lfrac{T_\text{eff}}{T}\right)^4
	\left(\ex^{\left(\lfrac{T}{T_\text{eff}}\right)x}+1\right)
	}
	{\ex^{\left(\lfrac{T}{T_\text{eff}}\right)x}-1}
	\right\rangle
	f
\right]
,
\end{equation}
where, for any function $q(v,\theta),$
\begin{equation}
\langle q\rangle
\equiv
\int_1^\infty\!\dd\gamma\, J(\gamma)\int_0^\pp\!\frac{\dd\theta \,\sin(\theta)}{2}\,q(v(\gamma),\theta)
\end{equation}
takes the appropriately\hyph{}weighted average over all relativistic gamma factors $\gamma$ and angles $\theta.$

\begin{figure}
	\centering
	\includegraphics[width=1.\linewidth]{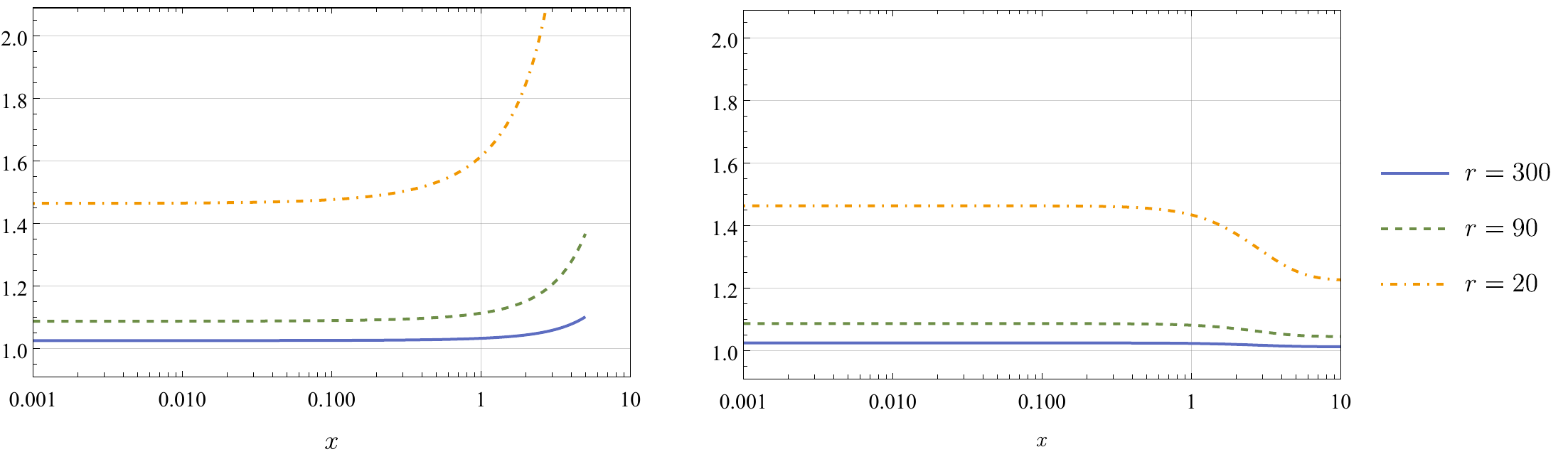}
	\caption{The ratio of angle\hyph{}bracket coefficients from the relativistic equation \eref{eq:f-rel} over the corresponding non\hyph{}relativistic coefficients from \eref{eq:f}. The plots are (\emph{left}) for the first angle\hyph{}bracket coefficient, associated with the $f$\hyph{}independent term, and (\emph{right}) for the second angle\hyph{}bracket coefficient, associated with the $f$\hyph{}dependent term.}
	\label{fig:rel}
\end{figure}

Equations~\eqref{eq:M-J} and~\eqref{eq:eff-T} enable calculation~\cite{num} of the two angle\hyph{}bracketed coefficients on the right\hyph{}hand side of \eref{eq:f-rel}.  Figure~\ref{fig:rel} 
 compares these with \eref{eq:f}'s corresponding non\hyph{}relativistic coefficients, and in interpreting the figure it should be noted that, for $x\gtrsim 1,$  the coefficient for the $f$\hyph{}independent term is exponentially suppressed relative to that for the $f$\hyph{}dependent term.  However, for $0.1\lesssim x \lesssim 1,$ relativistic effects increase the coefficient of the $f$\hyph{}independent term in \eref{eq:f-rel} more than the  $f$\hyph{}dependent term, and so tend to increase freeze\hyph{}out values.  A rough calculation of the effect on \dmwel mass can be made using the data in figures~\ref{fig:fignonrel} and \ref{fig:rel}, and this suggests that, for $\alpha_1=\num{e-2},$ the mass is increased by somewhere between $30$ and $60$~percent, equivalent to $0.1$ to $0.2$ dex, while, for $\alpha_1=\num{e-3},$ the mass increase is around $10$~percent.  The effects on emission\hyph{}absorption processes, important in sections~\ref{sec:cosmological-constraints} and~\ref{sec:detection-of-dmwel}, may well be less, as emissions and absorptions per \dmwel particle will increase, while the particle number density decreases. The conclusion is that condition (1) can be assumed to be satisfied for $\alpha_1\le\num{e-2},$ albeit possibly somewhat marginally right at the upper end of that range.  In any case, such corrections will not affect the overall\dash order of magnitude\dash  results in this paper, and will be disregarded.\\

\begin{figure}
	\centering
	\includegraphics[width=0.7\linewidth]{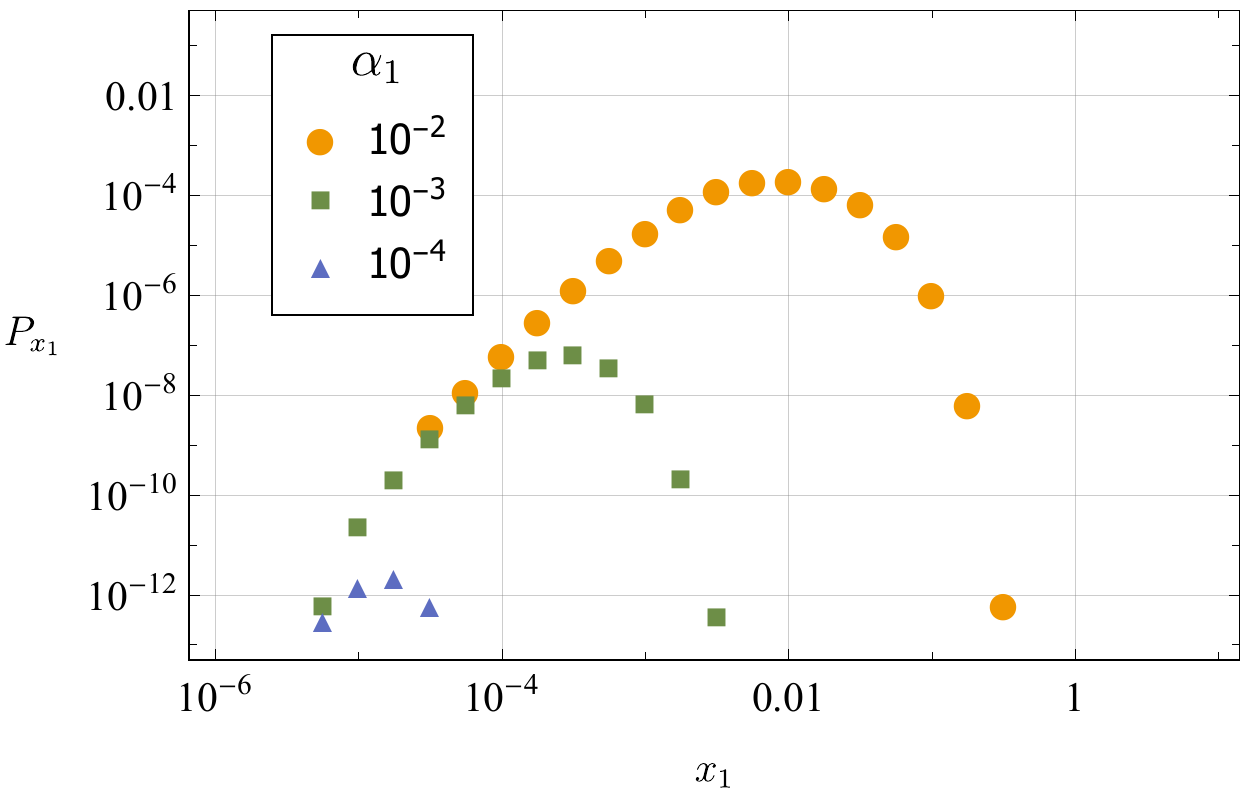}
	\caption{The probability $P_{x_1}$ as described in \eref{eq:P_x_1}, for selected $\alpha_1$ and $x_1.$}
	\label{fig:figenergycond}
\end{figure}

Now consider condition (2) to verify that for a stimulated emission or absorption the photon energy $E_\gamma$ is essentially the same as the level energy $E.$  When a stationary \dmwel particle absorbs a photon into a level of energy $E,$ conservation of energy and momentum implies that
\begin{equation}
E_\gamma
=
E
\left(1+\frac{E}{2 m}\right)
,
\end{equation} 
whilst stimulated emission from a stationary \dmwel particle\footnote{Noting that coherence of stimulated emission implies the emitted photon has the same energy and momentum as the stimulating photon.}  yields a photon of energy
\begin{equation}
E_\gamma
=
E
\left(1+\frac{E}{2\left(m+E\right)}\right)
.
\end{equation}
Absorption therefore implies a slightly tighter condition than simulated emission on $\lfrac{E}{m}$ in order to ensure $E_\gamma\approx E.$ Considering eqs.~\eqref{eq:ems-abs} and~\eqref{eq:f}, the combined rate of either an absorption or stimulated emission is given by
\begin{equation} \label{eq:combined}
\d{r_\text{c}}{x}
=
\frac{\alpha}
{x^3\left(\ex^{x}-1\right)}
.
\end{equation}
Figure~\ref{fig:figenergycond} shows the energy\hyph{}weighted  probability that an absorption or stimulated emission from a \dmwel particle is from a level $j$ that exceeds the arbitrary tolerance of $\lfrac{E_j}{2 m}\ge 0.01,$
\begin{equation}	\label{eq:P_x_1}
P_{x_1}
=
\left(\sum_{\left(\lfrac{E_j}{2 m}\right)\ge 0.01}
\!\!E_j\d{r_{\text{c}\,j}}{x_1}\right)
\bigg/
\left(\sum_{j\ge1}
E_j\d{r_{\text{c}\,j}}{x_1}\right)
.
\end{equation} 
Figure~\ref{fig:figenergycond} indicates that such absorptions or stimulated emissions, with $E_\gamma\not\approx E,$ are of minimal significance.
\\

It remains to check that the photon temperature is not significantly affected by \dmwel particles, at least other than for very high temperatures. From \figref{fig:f}, for large $\alpha_j$ (say $\alpha_j\gtrsim 100$), a useful approximation to $f_j$ is that it is ${1}/{2}$ for $x_j\lesssim 1$ and $0$ for $x_j\gtrsim 1$ (as an aside, but also as an accuracy check, this approximation can be used to explain the coincidence of low $x_1$ points for differing $\alpha_1$ in \figref{fig:fignonrel}).    The high\hyph{}$j$th level pairs  will emit, at around $x\sim 1,$ a photon energy per physical unit volume
\begin{equation}
\mathscr{E}_j
=
2 E_j n_0 \frac{E_j^3}{T_0^3}
.
\end{equation}
The transitioning level\hyph{}number $j(T)$ at photon temperature $T$ is given by $j(T)=\sqrt{\lfrac{T}{E_1}},$ and this gives an expression for the rate of change of radiation energy density with respect to temperature which includes the standard term $\lfrac{4\rho_\star}{T},$ and an additional term for the \dmwel effects, 
\begin{equation}
\d{\rho_\star}{T}
=
\frac{4\rho_\star}{T}
-
\frac{\mathscr{E}_j}{g_\gamma\sqrt{T E_1}}
,
\end{equation}
where $g_\gamma$ is the number of particle degrees of freedom in thermal equilibrium with photons, which is assumed to be constant at the maximum standard model value of $g_\gamma = {427}/{4}$~\cite{2018PhRvD..98c0001T}. This can be solved analytically~\cite{num} to give
\begin{mla}	\label{eq:rho-star}
	\frac{\rho_\star(T)}{\SI{e36}{GeV.cm^{-3}}}
	&=
	\left[\num{5e6}
	+
	\num{1e3}
	\refe^\infrac{-1}{2}
	\left(\frac{n_0}{n_{0\,\bullet}}\right)
	\right]\left(\frac{T}{\si{GeV}}\right)^4
	\nl
	-
	\num{8e1}
	\refe^\infrac{-1}{2}
	\left(\frac{n_0}{n_{0\,\bullet}}\right)\left(\frac{T}{\si{GeV}}\right)^\infrac{9}{2}
	,
\end{mla} 
with $n_{0\,\bullet}\equiv n_0|_{ \alpha_1=\num{e-3},\,E_1=\refequant}.$ Note, from \figref{fig:mfo-ndm0}~(right), that $n_{0\,\bullet}$ is, to order of magnitude, the biggest \dmwel number density compatible with \sref{sec:cmb-anisotropies}'s CMB anisotropy constraints.  This implies the second coefficient of $T^4$ in \eref{eq:rho-star} is always negligible.

From \eref{eq:rho-star}, it can be found that for 
\begin{equation}
T
\le
\refe \left(\frac{n_0}{n_{0\,\bullet}}\right)^{-2}
\SI{3e5}{GeV}
,
\end{equation}
corresponding to
\begin{equation}
j(T)
\le
\left(\frac{n_0}{n_{0\,\bullet}}\right)^{-1}
\num{4e6}
,
\end{equation}
the radiation density is no more than one percent different from its standard value~\cite{num}.  This justifies the assumption mentioned before \eref{eq:e-a-rules} that the maximum value of $j$ can be taken to be very large, and still allow the excitation equation, \eref{eq:time-deriv}, to be valid.  From \eref{eq:rho-star}, it can also be seen that the thermodynamic temperature has an asymptotic maximum value of 
\begin{equation}
T_\text{asymp}
=
\refe \left(\frac{n_0}{n_{0\,\bullet}}\right)^{-2}
\SI{4e9}{GeV}
,
\end{equation}
when all, or almost all, the future radiation energy is stored in \dmwel occupation energy.\footnote{\emph{Almost all} because when the temperature is very close to $T_\text{asymp},$ $\alpha_j$ being large is no longer ensures that $f_j$ is $0$ for $x_j\gtrsim 1$ because the photon energy density has fallen so far that energy levels no longer become unoccupied at a sufficient rate.} Absent any other new physics, at such very early times, the joint \dmwel\hyph{}radiation temperature is effectively constant at this value.  It is tempting to associate this epoch  with an end to inflation somewhat different from the usual reheating scenario.

\section{Photodissociation of nuclei}
\label{sec:photodissociation-of-light-nuclei}

Along lines indicated in, for example, refs.~\cite{2019JHEP...01..074F,2018JCAP...11..032H}, the evolution of light nuclei is governed by 
\begin{equation}
\d{Y_A}{t}
=
\sum_i Y_i \int_0^\infty\!\dd E_\gamma\, \mathscr{N}_\gamma(E_\gamma) \sigma_{\gamma+i\to A}(E_\gamma)
-
Y_A
\sum_j \int_0^\infty\!\dd E_\gamma\, \mathscr{N}_\gamma(E_\gamma) \sigma_{\gamma+A\to j}(E_\gamma)
,
\end{equation}
where for a nuclide $A,$ $Y_A\equiv\lfrac{n_A}{n_\text{b}},$ and the $\sigma$ represent the indicated cross sections. The quantity $\mathscr{N}_\gamma(E_\gamma)\equiv\ld{n_\gamma}{E_\gamma}$  is the spectral number density per unit energy of photons (in practice at the relevant photon energies and background photon temperature, solely consisting of \emph{non\hyph{}thermal} photons). For any given energy of emitted photons, the spectral number density is composed of a delta function summand at the emission energy plus a continuous spectrum at lower energies resulting from the down\hyph{}scattering of photons (including via down\hyph{}scattering to electrons). References~\cite{2019JHEP...01..074F,2018JCAP...11..032H} recently made detailed calculations of spectral number density suitable for photons emitted at energies $E\lesssim\SI{100}{MeV},$ which 
are the most relevant energies for \dmwel with $E_1$ near to the lower limits derived in \sref{sec:cmb-anisotropies}.  As indicated in \rcite[fig.~2]{2019JHEP...01..074F}, older approaches are not well\hyph{}suited for these emission energies, and can under\hyph{} or over\hyph{}estimate spectra by a factor of several. 
The calculations for this paper~\cite{num} closely follow the approach of \rcite{2019JHEP...01..074F}, and have been checked against results shown in that paper.  The cross-sections used are taken from \rcite{2003PhRvD..67j3521C}, except for $\isotope[7]{Li}+\gamma\to \isotope{T}+\isotope[4]{He}$ and $\isotope[7]{Li}+\gamma\to \isotope[3]{He}+\isotope[4]{He}$, where cross-sections are taken from \rcite{2014PhRvD..90h3519I}.

The effect of \dmwel emissions on light nuclides can be quantified via
\begin{equation} \label{eq:photodiss-quant}
\delta_A=\frac{\abs*{\Delta Y_A}}{O_A}
,\end{equation}
where $\Delta Y_A$ is the change in $Y_A$ arising from \dmwel emissions compared with the standard theoretical base\hyph{}case without these, and $O_A$ is the one\hyph{}sigma error estimate in the observed value of $Y_A.$ 
 \begin{table}
	\centering 
	\begin{tabular}{ccccc}
		\toprule
		Nucleus, $A$& \isotope{D} & \isotope[3]{He} & \isotope[4]{He} & \isotope{Li}  \\
		Maximum $\delta_A$ & $\num{6e-2}$ & $\num{6e-6}$ & $\num{2e-9}$ &$\num{3e-3}$  \\ 
		\bottomrule
	\end{tabular}
	\caption{The effect of \dmwel emissions on the abundance of light nuclei, as characterised in \eref{eq:photodiss-quant}. The values of $\delta_A$ are the maximums within the range ${10^{-4}\le\alpha_1\le 10^{-2}},$ consistent with the values of $E_1$ allowed by CMB anisotropies, which were derived in \sref{sec:cmb-anisotropies}.
	}
	\label{tab:photodiss}
\end{table} Recent values for the theoretical and observational abundance of nuclei can be found in \rcite{2018PhR...754....1P}.  Calculations confirm that $\delta_A$ always declines with increasing $E_1,$ at least once $E_1$ has increased a little above the minimum~\cite{num}.  As shown in Table~\ref{tab:photodiss}, for $10^{-4}\le\alpha_1\le 10^{-2},$ and $E_1$ allowed by CMB anisotropies,  $\delta_A$ is very small for each of the potentially observable nuclides, $A=\isotope{D},\,\isotope[3]{He},\,\isotope[4]{He},\,\isotope{Li}.$  (Here, \isotope{Li} includes both \isotope[6]{Li} and \isotope[7]{Li}, as well as  \isotope[7]{Be} which rapidly decays to lithium. Similarly \isotope{D} includes tritium, which rapidly decays to deuterium.) So, as claimed, the effects of \dmwel emissions on nuclear abundances are negligible, at least for such $\alpha_1.$ 


\addcontentsline{toc}{section}{\quad\ References}
\bibliographystyle{JHEP}
\bibliography{DMwEL_biblio}
\end{document}